\documentclass[a4paper,11pt]{article}

\usepackage{amssymb,amsmath,latexsym,amsthm}

\usepackage{mathabx}
\usepackage{xcolor}  
\usepackage{url}
\usepackage{hyperref}
\usepackage{cleveref}
\usepackage{enumerate}
\usepackage[square,sort,comma,numbers]{natbib}
\usepackage{dsfont, esint}
\usepackage{diagbox}
 \usepackage{geometry}
\usepackage{graphicx}
\usepackage{colortbl}
\usepackage{bbm}
\usepackage{mathtools}
\usepackage[title]{appendix}
\usepackage{cancel}

\usepackage{endnotes}

\renewcommand{\textbf}[1]{\begingroup\bfseries\mathversion{bold}#1\endgroup}

\usepackage{breqn}

\usepackage{fancyhdr}
\usepackage{pstricks,pst-plot,pst-node,pstricks-add}
\usepackage{auto-pst-pdf}

\newtheorem{thma}{Theorem}

\newtheorem{thm}{Theorem}[section]

\newtheorem{corollary}[thm]{Corollary}
\newtheorem{prop}[thm]{Proposition}

\newtheorem{lemma}[thm]{Lemma}

\theoremstyle{definition}
\newtheorem{rmk}[thm]{Remark}

\newcommand{\R}{\mathbb R}

\newcommand{\Z}{\mathbb Z}
\newcommand{\N}{\mathbb N}
\newcommand{\C}{\mathbb C}

\numberwithin{equation}{section}

\def\XXint#1#2#3{{\setbox0=\hbox{$#1{#2#3}{\int}$}
    \vcenter{\hbox{$#2#3$}}\kern-.5\wd0}}

\allowdisplaybreaks

\makeatletter
\def\blfootnote{\xdef\@thefnmark{}\@footnotetext}
\newcommand{\subjclass}[2][2020]{%
  \let\@oldtitle\@title%
  \gdef\@title{\@oldtitle\footnotetext{#1 \emph{Mathematics subject classification.} #2}}%
}
\makeatother

\date{date}
\let\underbrace\LaTeXunderbrace
\begin{document}

\title{Stability and robustness of mathematical quasicrystals under statistical convergence}
\author{Rodolfo Viera}

\newcommand{\Addresses}{{
  \bigskip
  \footnotesize
\textsc{Facultad de Ingeniería y Arquitectura, Universidad Central de Chile, Avda. Francisco de Aguirre 0405, La Serena, Chile.}\par\nopagebreak
\textit{E-mail address: }\texttt{rvieraq.docente@ucen.cl; r.vieraq@gmail.com}
}}

\date{\today}
\maketitle
\begin{abstract}
   In this work we address the stability and robustness of uniformly discrete point sets in Euclidean spaces. Firstly, we prove that if a sequence of point configurations contained in $\R^d$ is uniformly diffractive, converges rapidly enough to a discrete set $X$ in $\R^d$, and their diffraction measures $\widehat{\gamma_{X_n}}$ are asymptotically orthogonal with respect to the Lebesgue measure in $\R^d$, then $X$ is necessarily a quasicrystal. The convergence is addressed for a distance that quantifies the statistical closeness between two uniformly discrete point sets in $\R^d$. Secondly, motivated by their applications in the diffraction theory of quasicrystals, we establish the continuity of the Fourier Transform of quasicrystals in this topology. This continuity result, in turn, allows us to rigorously demonstrate that well-known robustness properties of quasicrystals against random errors remain stable under the statistical convergence considered. Some applications for rapidly-solidified quasicrystals are highlighted. 
\end{abstract}

\maketitle

\section{Introduction}

 Since the remarkable discovery of quasicrystals by Shechtman and his team in the 1980s, researchers in both physics and mathematics have made numerous efforts to understand their geometry and structural properties. These solid materials exhibit sharp peaks in their diffraction patterns, indicating long-range order, but their geometry witnesses rotational symmetries forbidden in classical crystals, due to the renowned Crystallographic Restriction Theorem; see the seminal work \cite{quasi}, also \cite{crysres1, crysres2}, and the books \cite{aper, BG2} for further details. A. Hof in \cite{Hof1} appears to have been the first to establish a rigorous mathematical framework for explaining the physicist's diffraction experiments and models. His work on mathematical diffraction and the study of quasicrystals have motivated further research from many branches of mathematics, such as Dynamical Systems \cite{aujogue, repquasi, autolim}, Mathematical Physics and Probability Theory \cite{Hof2, almostsure3} and Fourier Analysis \cite{contF, FourierMeyer}; see also \cite{probquasi} for an Oberwolfach report and \cite{aper, BG2} that count some of these efforts. It is worth noting that the mathematics of the aperiodic order have been studied by mathematicians even before the laboratory discovery of quasicrystals; here, we mention (though not exhaustively) some notable works, such as the construction of aperiodic tilings of the plane due to R. Penrose \cite{penrose}, the study of Almost periodic functions by H. Bohr \cite{almostperbohr} that motivated the so-called cut-and-project schemes to construct quasicrystals, and the relation between aperiodic structures and number theory \cite{meyer1}. These motivating discoveries remain a source of interesting mathematics, including the recent construction of the aperiodic monotile \cite{apmono} and subsequent studies of its diffractive properties \cite{Soc} and dynamics \cite{hatdyn}.
 \medskip

A major problem is to understand how {\em stable} and {\em robust} quasicrystals are. The notion of {\em stability} we adopt in this work is what properties of quasicrystals (or another point configuration) are preserved under (a suitable) convergence, and our goal is to prove that, under certain assumptions, such as a rate of convergence and an {\em equi-diffractive} hypothesis, the "quasicrystallinity" of a sequence of point configurations is preserved by the convergence we are considering; see Theorem \ref{thm: A} for a more detailed statement of our result. 
\medskip

Regarding the {\em robustness} of quasicrystals, in this work we mainly address the continuity of certain operators of interest in aperiodic order, specifically the Fourier Transform of a quasicrystal \cite{almostsure1, almostsure2}; this corresponds to Theorem \ref{thm: B} below. Additionally, we aim to prove that well-known robustness properties of quasicrystals against random errors remain stable under the convergence considered in this work. 
\medskip
 
We now proceed to state our results precisely. We say that a closed subset $X$ of $\R^d$ is {\em locally finite}, if for every bounded subset $B$ of $\R^d$ it holds that $\# X\cap B<\infty$. A locally finite point set $X$ in $\R^d$ is called {\em uniformly discrete}, if there is $r_0>0$ such that every ball in $\R^d$ of radius $r_0$ contains at most one point of $X$; the largest $r_0$ satisfying the above is called the {\em separation radius} of $X$. We denote the class of uniformly discrete subsets of $\R^d$ by $\mathcal{UD}_d$. For a fixed $r_0>0$, we denote by $\mathcal{UD}_{r_0,d}$ the collection of all the uniformly discrete point sets in $\R^d$ with separation radius bounded from below by $r_0$. This space appears suitable for modeling solid materials because there exists a minimal (physical) point distance, namely the Planck length $\hbar$\footnote{It is not our goal to delve into the quantum world.  What we intend to mention is that the (deterministic) inter-atomic distance in solid materials we are interested in, like (quasi)crystals,  cannot be as close as we want, or, in other words, their distances are (extremely more) larger than the Planck length.}; see \cite{calmet2007, calmet2005minimum} and the references therein. Thus, in our context, this makes it interesting to study point configurations as elements of $\mathcal{UD}_{\hbar,d}$.
 \medskip

In what follows, given a locally finite set $X$ in $\R^d$, we write,
 $X^{(L)}:=X\cap B_L$, where $B_L$ stands for the $d$-dimensional ball centered at the origin with radius $L$. The {\em autocorrelation measure} of a locally finite set $X$ in $\R^d$ is defined by the following limit (if it exists):
 \[
 \gamma_X=\lim_{L\to\infty}\frac{1}{L^d}\sum_{p,q\in X^{(L)}}\delta_{p-q};
 \]
 the measure $\gamma_X$ captures the correlations between points in $X$. The {\em diffraction measure of $X$} is the Fourier Transform of the autocorrelation measure, i.e., $\widehat{\gamma_X}$. We say that a locally finite subset $X\subset\R^d$ is a {\em mathematical quasicrystal} if the diffraction measure $\widehat{\gamma_X}$ is orthogonal with respect to the Lebesgue measure in $\R^d$. The careful reader should note that this definition does not require $X$ to be uniformly dense, nor that $\widehat{\gamma_{X}}$ is purely atomic. We adopt the terminology {\em mathematical quasicrystal} because it encompasses a broader class of point configurations with some kind of order, such as the visible points of the square lattice \cite{vis}, and uniformly discrete sets with a singular continuous diffraction measure \cite{sincon}.
\medskip

A nice way to establish dissimilarities between two measures $\mu,\nu$ is through the so-called (localized) {\em Hellinger density} $h(\mu,\nu)$. In this Introduction it is not necessary to define it precisely, but we only need to say is that two measures $\mu$ and $\nu$ are orthogonal if and only if $h(\mu,\nu)$ vanishes. We refer the reader to Section \ref{ssec: specstab} for further details. Finally, we say that a family $(X_{i})_{i\in I}$ of diffractive sets in $\mathcal{UD}_{r_0,d}$ is {\em uniformly diffractive} if there exists a Radon measure $\mu$ in $\R^d$ such that
 \[
 (\forall i\in I),\quad \widehat{\gamma_{X_i}}\ll\mu.
 \]
  
Let us now describe the convergence of locally finite point configurations we are interested in. Roughly speaking, we say that $X,Y\in\mathcal{UD}_{r_0,d}$ are at {\em statistical distance at most $\epsilon$} if for any ball centered at the origin, the set of points where they are $\epsilon$-apart is at most $\epsilon$ times a power of the volume of the ball. Specifically, for $X,Y\in\mathcal{UD}_{r_0,d}$ and $0<\alpha\leq 1$, we define
\begin{equation}\label{eq: statGHintro}
\rho_{\mathsf{stat},\alpha}(X,Y)=\min\left\{\inf\left\{\epsilon>0:\ \sup_{L>0}\frac{\#A_{X,Y}^{(L,\epsilon)}\cup A_{Y,X}^{(L,\epsilon)}}{L^{\alpha d}}<\epsilon\right\},\frac{r_0}{2}\right\},
\end{equation}

where $A_{X,Y}^{(L,\epsilon)}:=\{p\in X^{(L)}:\ d(p,Y^{(L)})\geq\epsilon\}$; in the case $\alpha=1$, we simply write $\rho_{\mathsf{stat},1}=\rho_{\mathsf{stat}}$. Later we will see that \eqref{eq: statGHintro} defines a complete metric space $(\mathcal{UD}_{r_0,d},\rho_{\mathsf{stat},\alpha})$ (see Lemmas \ref{lem: statdis} and \ref{lem: statcom} below), which we call the {\em statistical Hausdorff distance} between $X$ and $Y$. The definition \eqref{eq: statGHintro} provides a whole family of metrics $\rho_{\mathsf{stat},\alpha}$, where $\alpha\in (0,1]$ measures the admissible rates of dissimilarities between two uniformly discrete sets in $\mathcal{UD}_{r_0,d}$; in addition we have that $\rho_{\mathsf{stat},\alpha}(X,Y)<\rho_{\mathsf{stat},\beta}(X,Y)$ whenever $\beta<\alpha$. Observe that these types of distances are natural in the modelling of quasicrystals since distortion with respect to ideal atomic positions have been observed on them \cite{disquasi}. 
\medskip

Having this setting in hand, our first main Theorem then reads as follows.

\begin{thma}\label{thm: A}
    Let $r_0>0$ and $d\in\N$ be fixed. Let $X\in\mathcal{UD}_{r_0,d}$ and $(X_n)_{n\geq 1}\subset\mathcal{UD}_{r_0,d}$ be a sequence of diffractive point sets such that
    \begin{equation}\label{eq: thma}
        \rho_{\mathsf{stat},\frac{1}{2}}(X_n,X)\longrightarrow 0,\quad\text{when }n\text{ goes to }\infty.
    \end{equation}
    Then $X$ is diffractive too. In addition, if the $X_n$'s are uniformly diffractive and satisfy that
    \begin{equation}\label{eq: thmaort}
    h(\widehat{\gamma_{X_n}},\mathsf{Leb}_d)\xrightarrow[n\to\infty]{} 0\quad\text{vaguely},
    \end{equation}
    
    then $X$ is a quasicrystal.
\end{thma}

The first statement of Theorem \ref{thm: A} establishes that under statistical convergence at rate $\rho_{\mathsf{stat},1/2}\to 0$, diffractivity passes to the limit; here, the choice of $\alpha=1/2$ in \eqref{eq: thma} is because the proof of stability of autocorrelations needs error terms to disappear after normalization by $L^d$ while still allowing the bad set to be sparse enough to control pair correlations. In other words, a mismatch set of size $o(L^{d/2})$ is small enough that its contribution to double sums over pairs, which are of size $O(L^d)$ remains negligible in the limit. Following this discussion, the next question arises naturally from the statement of Theorem \ref{thm: A}: {\em under which conditions over the admissible $X_n$'s we have that condition \eqref{eq: thma} can be replaced by $\rho_{\mathsf{stat},\alpha}(X_n,X)\to 0$ , for some $\alpha\in\left(0,\frac{1}{2}\right)$?} We believe that this could be the case for $X_n$ being (almost) repetitive according \cite{repquasi, almrep, almostlin}, with constant $\alpha$ probably depending on the (almost-)repetitivity function of $X$, but this still remains open.
\medskip

The second part of Theorem \ref{thm: A} claims that under uniform diffractivity plus asymptotic orthogonality of the diffraction measures of the $X_n$'s to the Lebesgue measure, the limit must be a quasicrystal.
\medskip

Theorem \ref{thm: A} offers an idealized metric/spectral framework for the limiting quasicrystalline order observed in rapid solidification snapshot-by-snapshot: if a sequence of increasingly accurate snapshots of the solidification process has diffraction that is uniformly controlled and becomes orthogonal to Lebesgue measure in the limit, then the limiting structure must be a quasicrystal. In this interpretation, $\alpha$ can then be viewed as encoding how well quasicrystalline order emerges at each snapshot of the process; specifically, we have that values of $\alpha$ close to $1$ allow more incomplete matching at large balls with respect to the limit quasicrystal, while smaller values of $\alpha$ correspond to cleaner quasiperiodic organization. The diffuse or mixed diffraction seen in the intermediate stages corresponds to the approximating $X_n$, while the theorem says that once the statistical errors are sparse enough and the spectral behaviour is asymptotically orthogonal with respect to Lebesgue, the final structure retains quasicrystalline order rather than reverting to a disordered one. In view of this, we expect Theorem \ref{thm: A} to offer an idealized metric-spectral framework for the snapshot-by-snapshot analysis of experimental configurations associated with rapidly solidified quasicrystals observed in X-ray diffraction experiments \cite{quasi, rapsol1, rapsol2, rapsol3}. 
\medskip 

Another object we are interested in, due to its connection with the diffraction theory of quasicrystals, is the Fourier Transform of a uniformly discrete point configuration. For a locally finite point set $X\subset\R^d$, define its {\em Fourier Transform} by means of the formula:
\begin{equation}\label{eq: FTX}
   \mathcal{F}_X=\lim_{L\to\infty}\frac{\widehat{\delta_{X^{(L)}}}}{L^d},\footnote{Although we have the identification $X\leftrightarrow\delta_X$, $\mathcal{F}_X$ must not be confused with the Fourier Transform of the Dirac comb $\delta_X$ given by
   \[
  (\forall f\in\mathcal{S}(\R^d)),\quad \widehat{\delta_X}(f)=\delta_X(\widehat{f})=\sum_{p\in X}\widehat{f}(p),
   \]
   since the latter one is not necessarily a measure.}
\end{equation}
where the limit is taken in the vague sense. The limit in \eqref{eq: FTX} could exist or not, but it can be shown that the family $\frac{\widehat{\delta_{X^{(L)}}}}{L^d},\ L>0$, is precompact in the vague topology (see Lemma \ref{lem: FXmeas} below). Observe that, in the case when $X$ is an $d$-dimensional lattice, we recover the known averaged exponential sum formula:
\begin{equation}\label{eq: FXlatt}
\mathcal{F}_X(x)=\lim_{L\to\infty}\frac{1}{L^d}\sum_{p\in X^{(L)}}e^{-2\pi i\langle p,x\rangle}=\left\{\begin{array}{lcc}
V_d & \text{if} & x\in X^*\\
0 & \text{if} & x\not\in X^*\end{array}\right.=V_d\delta_{X^*}(x),
\end{equation}
where $X^*$ is the dual lattice of $X$ and $V_d$ denotes the volume of the unitary $d$-dimensional ball. 
\medskip

Even if one is not aware of the formula \eqref{eq: FXlatt} when $X$ is a lattice, we can show that $\mathcal{F}_X$ is not trivial whenever it exists. Indeed, if $f\in C_c(\R^d)$, then by definition of $\mathcal{F}_X$ we get:
\[
\mathcal{F}_X(f)=\lim_{L\to\infty}\frac{1}{L^d}\sum_{p\in X^{(L)}}\widehat{f}(p).
\]

The later limit does not (necessarily) vanish since $\widehat{f}$ cannot be compactly supported because the uncertainty principle (see for instance \cite{uncerprin}); hence the sum $\sum_{p\in X^{(L)}}\widehat{f}(p)$ depends heavily on $L$ which makes $\mathcal{F}_X$ non-trivial.
\medskip

The relevance of establishing properties for the Fourier Transform of quasicrystals is motivated by the fact that the diffraction intensities of their diffraction measures can be computed in terms of $\mathcal{F}_{X}$; indeed, it can be shown that for quasicrystals arising from regular model sets, their diffraction intensities $I(k)$ are given by:
\[
(\forall k\in\mathsf{supp}(\widehat{\gamma_X})),\quad I(k)=\left|\mathcal{F}_X(k)\right|^2,
\]
where $\mathsf{supp}(\widehat{\gamma})=L^{\ocoasterisk}$ is the Fourier module of the cut-and-project scheme used to construct $X$; we refer the reader to \cite[Section 9.4]{aper} for further details.
\medskip

In view of the above, an interesting problem is to explore further properties of the operator $\mathcal{F}:X\mapsto\mathcal{F}_X$. Let us define $\mathcal{Q}_{r_0,d}$ being the set of all {\em Fourier transformable} $r_0$-uniformly discrete point configurations in $\R^d$, namely:
\[
\mathcal{Q}_{r_0,d}:=\{X\in\mathcal{UD}_{r_0,d}:\ \mathcal{F}_X\text{ exists}\};
\]

this is a non-empty subset of $\mathcal{UD}_{r_0,d}$ containing lattices \cite[Corollary 9.3]{aper} and quasicrystals originated by cut-and-project schemes \cite[Proposition 9.9]{aper}. Consider $\mathcal{F}$ as a map defined on $\mathcal{Q}_{r_0,d}$ taking values in the space of complex measures $\mathcal{M}(\R^d)$. In addition, consider $\mathcal{Q}_{r_0,d}$  and $\mathcal{M}(\R^d)$ endowed with the statistical Hausdorff distance $\rho_{\mathsf{stat}}$ and the vague topology, respectively. Thus, with this setting in mind, our second main result reads as follows.

\begin{thma}\label{thm: B}
$\mathcal{Q}_{r_0,d}$ is a closed subset of $(\mathcal{UD}_{r_0,d},\rho_{\mathsf{stat}})$. Additionally, the map $\mathcal{F}:\mathcal{Q}_{r_0,d}\to\mathcal{M}(\R^d)$ is continuous.
\end{thma}

Theorem \ref{thm: B} makes the space $(\mathcal{Q}_{r_0,d},\rho_{\mathsf{stat}})$ into a natural metric-theoretic setting for the study of aperiodic order. It is worth to remark that, besides the well-known continuity of the Fourier transform of square-integrable functions $\mathcal{F}:L^2(\R^d)\to L^2(\R^d)$, the continuity of the Fourier Transform of measures as a map $\mathcal{F}:\mathcal{M}^{\infty}(\R^d)\to \mathcal{S}'(\R^d)$ has been addressed in \cite{contF}, where $\mathcal{M}^{\infty}(\R^d)$ is the space of the translation bounded measures and $\mathcal{S}'(\R^d)$ is the space of tempered distributions. However, these results are not directly connected with ours.
\medskip

It should be noted that the continuity of the above-defined Fourier Transform is especially relevant for quasicrystals, where one often has approximate experimental data rather than exact ideal sets.
\medskip
Lastly, we provide two applications of Theorem \ref{thm: B}, regarding the robustness of certain operators defined on locally finite sets in $\R^d$. For this, let $\xi_p,\ p\in X$ be iid random vectors defined in the same probability space $(\Omega,\mathcal{T},\mathbb{P})$, with common law $\xi_p\sim\xi$. Consider the {\em random perturbation} of $X$ under $\xi$ given by:
\begin{equation}\label{eq: randper}
    X_{\xi}:=\{p+\xi_p:\ p\in X\}.
\end{equation}

So we can take the Fourier Transform of the realization $X_{\xi}$ as follows:
\[
\mathcal{F}_{X_{\xi}}:=\lim_{L\to\infty}\frac{1}{L^d}\widehat{\delta_{X_{\xi}^{(L)}}}.
\]
We are concerned with the problem of whether or not we can recover $\mathcal{F}_X$ almost surely from the observations of $\mathcal{F}_{X_{\xi}}$. Let us be precise about what we mean by {\em recovering}; we say that $X$ is {\em Fourier-recoverable from its random perturbations} (or just {\em Fourier-recoverable}) if there exists a (deterministic) continuous map $\Phi: \mathcal{M}(\R^d)\to \mathcal{M}(\R^d)$ so that 

\begin{equation}\label{eq: recovFT1}
    \Phi(\mathcal{F}_{X_{\xi}})=\mathcal{F}_{X}\quad\text{almost surely};
\end{equation}

we call the map $\Phi$ in the above definition the {\em recovering map}. In \cite{almostsure1, almostsure2, almostsure3} it was shown, under certain control of the noise of the perturbations, that quasicrystals are Fourier-recoverable by a very explicit formula; more precisely, almost surely the following holds for every $x\in\R^d$:
\begin{equation}\label{eq: recovFT2}
\mathcal{F}_{X_{\xi}}(x)=\mathbb{E}\left(e^{-2\pi i\langle\xi,x\rangle}\right)\mathcal{F}_X(x),
\end{equation}
where $\mathcal{F}_X(x)$ in \eqref{eq: recovFT2} (respectively, $\mathcal{F}_{X_{\xi}}(x)$) is defined by
\begin{equation}
\mathcal{F}_X(x)=\lim_{L\to\infty}\frac{\widehat{\delta_{X^{(L)}}}(x)}{L^d}=\lim_{L\to\infty}\frac{1}{L^d}\sum_{p\in X^{(L)}}e^{-2\pi i\langle p,x\rangle},
\end{equation}
where $\widehat{\delta_{X^{(L)}}}(x),\ L>0$, are the Fourier-Stieltjes coefficients of $\delta_{X^{(L)}}$ (with a similar formula for $\mathcal{F}_{X_{\xi}}(x)$).
\medskip

By taking a look at the factorization formula \eqref{eq: recovFT2}, we observe the following facts: If $X\in\mathcal{Q}_{r_0,d}$ then $\mathcal{F}_{X_{\xi}}$ almost surely exists and, on the other hand, if $\mathbb{E}\left(e^{-2\pi i\xi\cdot x}\right)\neq 0$ (for instance, for a mixture of Gaussians), then $X$ is Fourier-recoverable from $X_{\xi}$ by means of:
\[
\mathbb{E}\left(e^{-2\pi i\langle\xi,x\rangle}\right)^{-1}\mathcal{F}_{X_{\xi}}(x)=\mathcal{F}_X(x);
\]
i.e., the recovering map is given by the multiplication operator $\Phi:f\mapsto \mathbb{E}\left(e^{-2\pi i\langle\xi,\cdot\rangle}\right)^{-1}f$ \footnote{After identifying a function $f\in L^2(\R^d)$ with the absolutely continuous measure with respect to the Lebesgue measure $f(x)dx$}. This was proved in \cite{almostsure3} for lattices, and for (aperiodic) quasicrystals in \cite{almostsure1,almostsure2}. Additionally, factorization formulae like \eqref{eq: recovFT2} are specially relevant since they make appear the modulating factor $\mathbb{E}(e^{-2\pi i\langle\cdot,\xi\rangle})$ that plays the same structural role as the {\em Debye-Waller factor} in crystallography \cite{debwall}.
\medskip

Given this, we are also interested in the stability of the Fourier-recoverability property under convergence. We address this in the case of {\em uniform recoverability}, namely, when the recovering map is uniform for every member of a family of uniformly discrete point sets. Specifically, given a continuous map $\Phi:\mathcal{M}(\R^d)\to \mathcal{M}(\R^d)$, we say that a family $(X_{\alpha})_{\alpha\in I}$ of uniformly discrete sets in $\R^d$, each one with its own family of iid random vectors $\xi_p^{(\alpha)},\ p\in X_{\alpha}$, is {\em $\Phi$-Fourier-recoverable} if for every $\alpha\in I$ we have that $X_{\alpha}$ is Fourier -recoverable with recovering function $\Phi$. With this setting in mind, we are in a position to state our third main result, which reads as follows.

\begin{thma}\label{thm: C}
    Let $r_0>0$ and $d\in\N$ be fixed. Let $X_n,\ n\in\N$, and $X$ be uniformly discrete sets in $\mathcal{Q}_{r_0,d}$ having asymptotic densities, and satisfying that $X_n\xrightarrow{\rho_{\mathsf{stat}}}X$. Let $\xi^{(n)}_q,\ q\in X_n$ and $\xi_p,\ p\in X$ be two sequences of iid random vectors defined in the same probability space $(\Omega,\mathcal{T},\mathbb{P})$, with common laws $\xi_p^{(n)}\sim\xi^{(n)},\ n\in\N,\ p\in X_n$, and $\xi_p\sim\xi,\ p\in X$. Also assume there are positive numbers $s,\epsilon0$ such that the following holds: 
    \begin{enumerate}
        \item $\xi_p^{(n)}$ is independent of $\xi_q$, provided $\|p-q\|<s$;
        \item $\mathbb{E}(\|\xi\|^{d+\epsilon})<\infty$.
    \end{enumerate}
    If $\xi^{(n)}\xrightarrow{\mathsf{law}}\xi$ and the $X_n$'s are uniformly Fourier-recoverable, then $X$ is Fourier-recoverable too.
\end{thma}

Broadly speaking, Theorem \ref{thm: C} says that if a quasicrystalline inverse problem is solvable uniformly along a statistically convergent family of approximations, then it remains solvable in the limit. It should be noted, as a consequence of the main results of \cite{almostsure1, almostsure2}, that if $(X_n)_{n\geq 1}\subset\mathcal{Q}_{r_0,d}$ and $X$ satisfy the hypotheses of Theorem \ref{thm: A}, and the noises $\xi_p^{(n)},\ n\in\N,\ p\in X_n$ and $\xi_p,\ p\in X$ satisfy the hypotheses of Theorem \ref{thm: C}, then $X$ is certainly Fourier recoverable. Thus, Theorem \ref{thm: C} becomes interesting since no convergence rate for $X_n\xrightarrow{\rho_{\mathsf{stat}}} X$ is required, unlike \eqref{eq: thma}.
\medskip

The motivating scheme for Theorem \ref{thm: C} is the following: suppose we have a sequence $X_n,\ n\in\N$ of inaccurate "snapshots" of a uniformly discrete set $X$ for which we would like to determine whether it is Fourier recoverable. Theorem \ref{thm: C} ensures that if the Fourier transforms of the observations $X_n$'s can be reconstructed through the same mechanism $\Phi$, and $X_n\to X$ statistically, then we can recover $\mathcal{F}_X$ through the limit of the $\mathcal{F}_{(X_n)_{\xi^{(n)}}}$.  

\subsection*{Sketch of proofs}

We split the proof of Theorem \ref{thm: A} in two parts: firstly, we show that if the $X_n$'s are diffractive, then $X$ is also diffractive and the sequence $(\gamma_{X_n})_{n\in\N}$ of the autocorrelations of the $X_n$'s converges (under a subsequence) vaguely to $\gamma_X$; we shall prove this in Proposition \ref{prop: stabdiff} below. Similar and independent results have been obtained in this direction in \cite{autolim} for the Chabauty-Fell convergence of Delone sets; there, the authors prove that for a uniquely ergodic Delone set $X$, its autocorrelation measure can be approximated by a sequence of autocorrelations of periodic approximations of $X$. Unique ergodicity is known to be equivalent to the condition that not only patches in $X$ appear repetitively, but also with a well-defined strictly positive frequency (see \cite{quasi} for further details). The strength of our result is that it does not require any information over the limit (except that it has the same separation radius as the members of the sequence), but only on the sequence $(X_n)_{n\geq 1}$ and their autocorrelation measures. Secondly, in Proposition \ref{prop: stabspec} we prove that if the sequence $(X_n)_{n\geq 1}$ is uniformly diffractive (see \eqref{eq: unifquasicrys}), then $h(\widehat{\gamma_{X_n}},\mathsf{Leb}_d)\to h(\widehat{\gamma_X},\mathsf{Leb}_d)$ in the vague sense, where $h$ is the localized Hellinger density. Therefore, under the hypothesis \eqref{eq: thmaort}, we conclude $X$ must be a quasicrystal since $h(\widehat{\gamma_X},\mathsf{Leb}_d)$ vanishes.
\medskip

Concerning Theorem \ref{thm: B}, to prove the continuity of the Fourier transform $\mathcal{F}:\mathcal{Q}_{r_0,d}\to\mathcal{M}(\R^d)$, given a sequence $X_n\xrightarrow{\rho_{\mathsf{stat}}}X$ and a compactly supported function $f:\R^d\to\C$, we need to show that $\mathcal{F}_{X_n}(f)\to\mathcal{F}_X(f)$. To achieve this, we prove that
\[
\left|\mathcal{F}_X(f)-\mathcal{F}_{X_n}(f)\right|\leq\text{(geometric error) }+\text{ (mismatch error)},
\]
and both terms in the right-hand side tend to 0. 
\medskip

Regarding Theorem \ref{thm: C}, a key step in proving it is the continuity of the Fourier Transform $\mathcal{F}$ of uniformly discrete point configurations given in Theorem \ref{thm: B}. The most technical step is to show, under the hypotheses of the noises $\xi^{(n)},\xi$ given in the statement of Theorem \ref{thm: C}, that $\mathcal{F}_{(X_n)_{\xi^{(n)}}}\to\mathcal{F}_{X_{\xi}}$; this corresponds to Proposition \ref{lemma: convFTper}. The proof of Theorem \ref{thm: C} is deferred to Appendix \ref{ssec: stabrob}. 
\subsection*{Notations}
Throughout this work, we denote by $B_L,\ L>0$, the $d$-dimensional Euclidean ball centered at the origin with radius $L$. The dot product $\R^d$ is denoted by $\langle\cdot,\cdot\rangle$. By $C_{r,K}$ we always denote a positive constant depending on the parameters $r,K$ (or even more parameters, which can be numbers, or subsets of $\R^d$). The {\em characteristic function} over a set $A\subset\R^d$ is denoted by $\chi_A$ and defined as:
\[
\chi_A(x)=\left\{\begin{array}{lcc}
   1  & \text{if} & x\in A \\
   0  & \text{if} & x\not\in A
\end{array}
\right.
\]

In the case $A=\{p\},$ we write $\delta_p:=\chi_{\{p\}}$, which is called the Dirac delta at $p\in\R^d$.
\medskip

For a locally finite set $X\subset\R^d$, we write
\[
X^{(L)}:=X\cap B_L.
\]
The Lebesgue measure in $\R^d$ is denoted by $\mathsf{Leb}_d$. We write $C_c(\R^d)$ for the set of all the continuous functions $f:\R^d\to\C$ for which their support $\mathsf{supp}(f):=\overline{\{x\in\R^d:\ f(x)\neq 0\}}$ is compact; as usual, we denote by $\|f\|_{\infty}$ the uniform norm.  
\medskip

Given $U,V$ be two closed subsets of $\R^d$, define its {\em Hausdorff distance} by

\[
d_H(U,V):=\inf\{\epsilon>0:\ U\subset B(V,\epsilon)\ \wedge\ V\subset B(U,\epsilon)\},
\]
where $B(A,\epsilon)$ is the $\epsilon$-neighborhood of $A\subset\R^d$. The {\em Hausdorff distance} between $X,Y\in\mathcal{UD}_{d}$ is defined by:
\[
\rho_{H}(X,Y)=\min\left\{\frac{1}{4},\widetilde{d_H}(X,Y)\right\},\quad\text{where }\widetilde{d_H}(X,Y):=\inf\{\epsilon>0: d_H(X^{(1/\epsilon)},Y^{(1/\epsilon)})\leq\epsilon\}.
\]

\section{Preliminaries on Measures and Diffraction Theory}\label{ssec:diffraction}
In this Section, we introduce the basics of mathematical diffraction theory used throughout this work; these ideas were initially established by A. Hof in \cite{Hof1} (see also \cite{aper}). Let $\mathcal{S}(\mathbb{R}^d)$ be the {\em Schwartz space} of rapidly decreasing $C^{\infty}$ (complex-valued) test functions. The {\em Fourier transform} of $f\in\mathcal{S}(\mathbb{R}^d)$ is defined by

\begin{equation*}
    \mathcal{F}(f)(x)=\widehat{f}(x):=\int f(\lambda)e^{-2\pi i\langle x,\lambda\rangle}d\lambda.
\end{equation*}

By a (complex) {\em measure} we mean a linear functional on the set of compactly supported functions $C_c(\mathbb{R}^d)$ satisfying the following: for every compact set $K\subset\mathbb{R}^d$ there is a positive constant $M_K$ such that

\begin{equation*}
    |\mu(f)|\leq M_K||f||_{\infty},
\end{equation*}

for every $f\in C_c(\mathbb{R}^d)$ supported on $K$, and where $||\cdot||_{\infty}$ denotes the supremum norm on $C_c(\mathbb{R}^d)$; the space of complex measures on $\R^d$ is denoted by $\mathcal{M}(\R^d)$. By the Riesz Representation Theorem, there is an equivalence between this definition of measure and the classical measure-theoretic concept of regular Radon measure. For a Radon measure $\mu$ on $\R^d$, there exists a largest open set $G$ for which $\mu(G)=0$; the complement $\R^d\setminus G$ is called the {\em support} of $\mu$, and denoted by $\mathsf{supp}(\mu)$.
\medskip

When a measure $\mu$ determines a tempered distribution $T_{\mu}(f)=\mu(f)$ (i.e a continuous linear functional defined on $\mathcal{S}(\mathbb{R}^d)$) via the formula

\begin{equation*}
    \forall f\in\mathcal{S}(\mathbb{R}^d),\qquad T_{\mu}(f):=\int f(x)d\mu(x),
\end{equation*}

then its Fourier transform is defined by $\widehat{\mu}(f):=\mu(\widehat{f})$.\\

The {\em total variation} $|\mu|$ of a measure $\mu$ is the smallest positive measure for which $|\mu(f)|\leq |\mu|(f)$ holds for every non-negative $f\in C_c(\mathbb{R}^d)$. A measure $\mu$ is said to be {\em translation bounded} if for every compact $K\subset\mathbb{R}^d$ there is a positive number $C_K$ such that $|\mu|(x+K)\leq C_K$, for every $x\in\mathbb{R}^d$. A family $(\mu_{\alpha})_{\alpha\in I}$ is called {\em uniformly (or equi-) translation bounded} if for every compact set $K\subset\R^d$, there exists a positive constant $C_K$ such that:

\begin{equation}\label{eq: uniftb}
   (\forall\alpha\in I),\quad \sup_{x\in\R^d}|\mu_{\alpha}|(x+K)\leq C_K. 
\end{equation}

Let $P_{\mu}:=\{x\in\mathbb{R}^d:\ \mu(\{x\})\neq 0\}$ be the set of atoms of $\mu$. We say that $\mu$ is {\em purely atomic} (or {\em pure point}) if it has atoms only, i.e, $\mu(A)=\sum_{x\in A\cap P_{\mu}}\mu(\{x\})$ for every Borel set $A$; moreover $\mu$ is called {\em continuous} if $\mu(\{x\})=0$ for every $x\in\mathbb{R}^d$. A measure $\mu$ is said to be {\em absolutely continuous} with respect to a measure $\nu$, if there exists $g\in L^{1}_{loc}(\nu)$ such that $\mu=g\nu$. Furthermore $\mu$ and $\nu$ are called {\em mutually singular} (which is denoted by $\mu\perp\nu$) if there exists a relatively compact Borel set $S\subset\mathbb{R}^d$ such that $|\mu|(S)=0$ and $|\nu|(\mathbb{R}^d\setminus S)=0$ (with obvious interpretation for the value of $|\mu|(A)$).\\

 For $f\in\mathcal{S}(\mathbb{R}^d)$, define $\widetilde{f}(x):=\overline{f(-x)}$, and for a measure $\mu$ we consider $\widetilde{\mu}(f):=\overline{\mu(\widetilde{f})}$. We say that $\mu$ is {\em positive definite} if $\mu(f*\widetilde{f})\geq 0$ for every $f\in C_c(\mathbb{R}^d)$. By Proposition 8.6 in \cite{aper}, every positive definite measure is Fourier transformable, and its Fourier transform is a positive translation-bounded measure.\\
 
 Recall that the {\em convolution} $\mu*\nu$ between two measures $\mu$ and $\nu$ is defined by 

\begin{equation*}
    \mu*\nu(f):=\displaystyle\int f(x+y)d\mu(x)d\nu(y);
\end{equation*}

If $\mu$ is translation bounded, and $\nu$ is finite, then the convolution $\mu*\nu$ is well-defined and is also a translation bounded measure. Moreover, whenever $\widehat{\mu}$ is a measure, one has the identity $\widehat{\mu*\nu}=\widehat{\mu}\widehat{\nu}$ (see Theorem 8.5 in \cite{aper}).\\

We say that a net of measures $(\mu_R)_{R>0}$ (resp. positive definite measures) {\em converges vaguely} to a measure $\mu$ if for every test function $f\in C_c(\mathbb{R}^d)$

\begin{equation*}
    \lim_{R\to\infty}\mu_R(f)=\mu(f).
\end{equation*}

Regarding a compactness criterion for measures with the vague topology, we use the following Proposition that can be found in \cite[Proposition 3.2]{contF}.

\begin{prop}\label{prop: compvag}
    Let $\mathcal{A}\subset\mathcal{M}(\R^d)$. Then the following are equivalent:
    \begin{enumerate}
        \item $\mathcal{A}$ is precompact.
        \item  $\mathcal{A}$ is vaguely bounded.
        \item For each compact set $K\subset\R^d$, the set $\{|\mu|(K):\ \mu\in\mathcal{A}\}$ is bounded.
    \end{enumerate}
\end{prop}
In particular, a family of uniformly translation-bounded measures satisfies 2 and hence is precompact. 


We denote by $\mu_{B_R}$ the restriction of $\mu$ over the ball $B_R$. The {\em autocorrelation measure of $\mu$} is defined by

\begin{equation}\label{eq: autoc}
    \gamma_{\mu}:=\lim_{R\to\infty}\frac{1}{R^d}(\mu_{B_R}*\widetilde{\mu_{B_R}}),
\end{equation}

whenever this limit exists. In this case, $\gamma_{\mu}$ is the limit of positive definite measures; thus, it is positive definite, and then it is Fourier transformable, and its Fourier transform is positive definite and translation bounded. The Fourier Transform $\widehat{\gamma_{\mu}}$ is called the {\em diffraction measure of $\mu$}.\\ 

If $\mu:=\sum_{p\in X}\mu(p)\delta_p$, where $\delta_p$ is the Dirac measure at $p\in X$, to deal with the diffraction measure of $\mu$ we need to prove that, in the sense of vague convergence of measures,

\begin{equation*}
    \lim_{R\to \infty} \frac{1}{R^d}\left|\sum_{p\in X\cap B_R}\mu(p)e^{-2\pi i \langle p,y\rangle}\right|^2 dy= \widehat{\gamma_{\mu}}
\end{equation*}

In more abstract terms, we are required to prove the following:
\begin{equation}\label{F-lim=lim-F}
    \lim_{R\to\infty}\mathcal F\left(\frac{1}{R^d} \mu_{B_R}\ast \widetilde{\mu_{B_R}}\right) =\mathcal F\left(\lim_{R\to\infty} \frac{1}{R^d}\mu_{B_R}\ast \widetilde{\mu_{B_R}}\right),
\end{equation}

where $\mathcal{F}$ denotes the Fourier transform. This follows from in \cite[Lemma 4.11.10]{BG2}, which states that if a net of positive definite measures converges weakly in the sense of measures, then their Fourier transforms converge weakly as well.
\section{Statistical closeness and convergence}\label{sec: statconv}

Henceforth, let $r_0>0$ and $d\in\N$ be fixed. As we discussed in the Introduction, we are concerned with the stability of properties defined over locally finite sets, where the convergence we consider witnesses the statistical closeness of such discrete sets. Define the {\em statistical Hausdorff distance} between $X,Y\in\mathcal{UD}_{r_0,d}$ by
\begin{equation}\label{eq: statGH}
    \rho_{\mathsf{stat}}(X,Y):=\min\left\{\widetilde{\rho_{\mathsf{stat}}}(X,Y),\frac{r_0}{2}\right\},
\end{equation}

where
\begin{equation}\label{eq: statGH2}
\widetilde{\rho_{\mathsf{stat}}}(X,Y):=\inf\left\{\epsilon>0:\ \sup_{L>0}\frac{\#A_{X,Y}^{(L,\epsilon)}\cup A_{Y,X}^{(L,\epsilon)}}{L^d}<\epsilon\right\},
\end{equation}
and $A_{X,Y}^{(L,\epsilon)}$ is the set of the points in $X^{(L)}$ that are at distance at least $\epsilon$ from $Y^{(L)}$, that is to say, 
\[
A_{X,Y}^{(L,\epsilon)}:=\left\{x\in X^{(L)}:\ d(x,Y^{(L)})\geq\epsilon\right\};
\]
observe that the sets $A_{X,Y}^{(L,\epsilon)}$ and $A_{Y,X}^{(L,\epsilon)}$ are disjoint, and thus it holds 
\[
\#A_{X,Y}^{(L,\epsilon)}\cup A_{Y,X}^{(L,\epsilon)}=\# A_{X,Y}^{(L,\epsilon)}
+\#A_{Y,X}^{(L,\epsilon)}.
\]
The minimum in the formula \eqref{eq: statGH} is taken just in order to obtain a well-defined metric, as we shall see in Lemma \ref{lem: statdis}. A simple remark regarding formulas \eqref{eq: statGH} and \eqref{eq: statGH2} is the following: let $0<\epsilon<\min\{r_0/2,1\}$ be given. Then:
\begin{equation}\label{eq: obsrhostat}
    \rho_{\mathsf{stat}}(X,Y)<\epsilon\quad\Rightarrow\quad d_{H}\left(X^{(\epsilon^{-1/d})},Y^{(\epsilon^{-1/d})}\right)<\epsilon,
\end{equation}
 where $d_H$ stands for the Hausdorff distance between compact subsets of $\R^d$. In particular, if $\rho_{\mathsf{stat}}(X_n,X)\to 0$, then $d_{H}(X_n^{(L)},X^{(L)})\to 0$ for any $L>0$. Now we prove first that $\rho_{\mathsf{stat}}:\mathcal{UD}_{r_0,d}\times\mathcal{UD}_{r_0,d}\to\R_{\geq 0}$ is actually a metric.

\begin{lemma}\label{lem: statdis}
    $(\mathcal{UD}_{r_0,d},\rho_{\mathsf{stat}})$ is a metric space.
\end{lemma}

\begin{proof}
   It is quite direct that $\rho_{\mathsf{stat}}(X,Y)=\rho_{\mathsf{stat}}(Y,X)$ and that $\rho_{\mathsf{stat}}(X,X)=0$. The fact that 
    \[
    \rho_{\mathsf{stat}}(X,Y)=0\quad\Rightarrow\quad X=Y
    \]
    follows from the property \eqref{eq: obsrhostat} by taking $\epsilon\to 0$. It only remains to prove the triangle inequality in order to obtain that $\rho_{\mathsf{stat}}$ defines a distance in $\mathcal{UD}_{r_0,d}$. 
    \medskip
    Let $X,Y,Z\in\mathcal{UD}_{r_0,d}$, and let
    \[
    \rho_{\mathsf{stat}}(X,Z)=\epsilon_1,\quad \rho_{\mathsf{stat}}(X,Y)=\epsilon_2,\quad \rho_{\mathsf{stat}}(Y,Z)=\epsilon_3; 
    \]
    moreover, assume that $\epsilon_1>\max\{\epsilon_2,\epsilon_3\}$ (if $\epsilon_1\leq\epsilon_2$ or $\epsilon_1\leq\epsilon_3$, we automatically get that $\epsilon_1\leq\epsilon_2+\epsilon_3$); also assume that $\max\{\epsilon_2,\epsilon_3\}<r_0/2$, otherwise, we immediately obtain that $\epsilon_1\leq\epsilon_2+\epsilon_3$ . We need to show that
    \begin{equation}\label{eq: triangstat}
    \sup_{L>0}\frac{\# A_{X,Z}^{(L,\epsilon_2'+\epsilon_3')}+\# A_{Z,X}^{(L,\epsilon_2'+\epsilon_3')}}{L^d}\leq\epsilon_2'+\epsilon_3',
    \end{equation}

    for any $\epsilon_2',\epsilon_2'\in (0,r_0/2)$ for which 
    \[
    \sup_{L>0}\frac{\# A_{X,Y}^{(L,\epsilon_2')}\cup A_{Y,X}^{(L,\epsilon_2')}}{L^d}<\epsilon_2'\quad\text{and}\quad \sup_{L>0}\frac{\# A_{Y,Z}^{(L,\epsilon_3')}\cup A_{Y,Z}^{(L,\epsilon_3')}}{L^d}<\epsilon_3';
    \]
    The conclusion follows by taking the corresponding infimum.
    \medskip
     
    To prove this, let $p\in A_{X,Z}^{(L,\epsilon_2'+\epsilon_3')}$ with $p\not\in A_{X,Y}^{(L,\epsilon_2')}$; in particular, there must exist $y_p\in Y^{(L)}$ such that
    \[
    \|p-y_p\|<\epsilon_2'.
    \]
    Thus, by triangle inequality, we have the following for every $z\in Z^{(L)}$:
    \[
    \epsilon_2'+\epsilon_3'\leq \|p-z\|\leq\|p-y_p\|+\|y_p-z\|<\epsilon_2'+\|y_p-z\|,
    \]
    and thus $\|z-y_p\|\geq \epsilon_3'$. In other words, we obtain that $y_p\in A_{Y,Z}^{(L,\epsilon_3')}$. Observe that the point $y_p$ must be unique; indeed, if $y_p, y_p'\in Y^{(L)}$ both satisfy that
    \[
    \|p-y_p\|<\epsilon_2'\quad\text{and}\quad \|p-y_p'\|<\epsilon_2',
    \]
    then, by triangle inequality and the definition of $\epsilon_2'$, we get that
    \[
    \|y_p-y_p'\|\leq \|y_p-p\|+\|p-y_p'\|<2\epsilon_2'<r_0,
    \]
    which is impossible since the separation radius of $Y$ is at least $r_0$. Hence, we have the assignation $A_{X,Z}^{(L,\epsilon_2'+\epsilon_3')}\setminus A_{X,Y}^{(L,\epsilon_2')}\ni p\mapsto y_p\in A_{Y,Z}^{(L,\epsilon_3')}$, which results to be one-to-one; otherwise, if $p,p'\in A_{X,Z}^{(L,\epsilon_2'+\epsilon_3')}\setminus A_{X,Y}^{(L,\epsilon_2')}$ are so that $y_p=y_p'$, then again by triangle inequality:
    \[
    \|p-p'\|\leq \|p-y_p\|+\|p'-y_p'\|<2\epsilon_2'<r_0.
    \]
    Thus, from this assignation, we get that 
    \[
    \#A_{X,Z}^{(L,\epsilon_2'+\epsilon_3')}\setminus A_{X,Y}^{(L,\epsilon_2')}\leq \#A_{Y,Z}^{(L,\epsilon_3')},
    \]
    which together the obvious inequality $\# A_{X,Z}^{(L,\epsilon_2'+\epsilon_3')}\cap A_{X,Y}^{(L,\epsilon_2')}\leq \# A_{X,Y}^{(L,\epsilon_2')}$, yields that
    \begin{equation}\label{eq: statdis1}
        \# A_{X,Z}^{(L,\epsilon_2'+\epsilon_3')}\leq \# A_{X,Y}^{(L,\epsilon_2')}+\# A_{Y,Z}^{(L,\epsilon_3')}.
    \end{equation}
    By an argument of symmetry, one can easily obtain that:
    \begin{equation}\label{eq: statdis2}
        \# A_{Z,X}^{(L,\epsilon_2'+\epsilon_3')}\leq \# A_{Z,Y}^{(L,\epsilon_3')}+\# A_{Y,X}^{(L,\epsilon_2')}.
    \end{equation}
    By putting together \eqref{eq: statdis1} and \eqref{eq: statdis2}, we get
    \[
    \# A_{X,Z}^{(L,\epsilon_2'+\epsilon_3')}+\# A_{Z,X}^{(L,\epsilon_2'+\epsilon_3')}\leq (\# A_{X,Y}^{(L,\epsilon_2')}+\# A_{Y,X}^{(L,\epsilon_2')})+ (\# A_{Y,Z}^{(L,\epsilon_3')}+\# A_{Z,Y}^{(L,\epsilon_3')});
    \]
    after dividing by $L^d$, and the definition of $\epsilon_2'$ and $\epsilon_3'$, the following holds:
    \[
    \frac{\# A_{X,Z}^{(L,\epsilon_2'+\epsilon_3')}+\# A_{Z,X}^{(L,\epsilon_2'+\epsilon_3')}}{L^d}\leq\epsilon_2'+\epsilon_3'.
    \]
This concludes the proof of the Lemma.
\end{proof}
 \begin{rmk}\label{rmk statdis}
 From the proof of Proposition \ref{lem: statdis}, it is worth noting that the averaging factor $L^d$ does not play any special role in the fact that $\rho_{\mathsf{stat}}$ is a distance. We could replace $L^d$ by, for instance $L^{\alpha d}$, where $0<\alpha\leq 1$, and still obtain a distance function $\rho_{\mathsf{stat},\alpha}$. The different values that one can consider in the power of $L^{(\cdot)}$ could be thought of as different rates of dissimilarity between two discrete sets. 
 \end{rmk}
\begin{lemma}\label{lem: statcom}
    $(\mathcal{UD}_{r_0,d},\rho_{\mathsf{stat}})$ is complete.
\end{lemma}
\begin{proof}
    Let $(X_n)_{n\in\N}\subset\mathcal{UD}_{r_0,d}$ be a Cauchy sequence with respect to $\rho_{\mathsf{stat}}$. By \eqref{eq: obsrhostat} we have that, for every $L>0$ there holds:
    \[
    d_{H}(X_n^{(L)}, X_m^{(L)})\to 0\quad\text{when }m,n\to\infty.
    \] 
    In particular, for each $k\in\N$ fixed, the sequence $(X_n^{(2^k)})_{n\in\N}$ converges to a finite point set $X^{(2^k)}\subset B_{2^k}$, for which 
    \[
    (\forall x,y\in X^{(2^k)}, x\neq y),\quad\|x-y\|\geq r_0;
    \]
    moreover, it holds that $X^{(2^k)}\subset X^{(2^{k+1})}$. Define $X$ being the union of the $X^{(2^k)}$'s, $k\in\N$, that is to say:
\[ X:=\bigcup_{k\in\N} X^{(2^k)}; \] 
    we claim that $X\in\mathcal{UD}_{r_0,d}$ and $X_n\xrightarrow{\rho_{\mathsf{stat}}}X$. 
    \medskip

    The fact that $X\in\mathcal{UD}_{r_0,d}$ follows directly from \eqref{eq: obsrhostat} and the definition of $X^{(2^k)}$. Now we prove that $\rho_{\mathsf{stat}}(X_n,X)\to 0$; to achieve this, observe that for each $k\in\N$, there exists $n_k\in\N$ such that
    \begin{equation}\label{eq: statcomp1}
        d_{H}(X_{n_k}^{(2^k)},X^{(2^k)})<2^{-k}\quad\text{and}\quad \rho_{\mathsf{stat}}(X_{n_k},X_{n_{k+1}})<2^{-(k+1)}.
    \end{equation}
    On the one hand, note that
    \begin{equation}\label{eq: statcomp2}
   d_{H}(X_{n_k}^{(2^k)}, X^{(2^k)})<2^{-k}\quad\Leftrightarrow\quad \# A_{X_{n_k},X}^{(2^k,2^{-k})}\cup A_{X,X_{n_k}}^{(2^k,2^{-k})}=0.
    \end{equation}
    On the other hand, from the proof of Lemma \ref{lem: statdis} (see inequality \eqref{eq: statdis2}), the following holds for every large-enough $k\in\N$:
    \begin{equation}\label{eq: statcomp3}
    \begin{split}
    \frac{\# A_{X_{n_{k}},X}^{(2^{k+1},2^{-k})}\cup  A_{X,X_{n_{k}}}^{(2^{k+1},2^{-k})}}{2^{(k+1)d}} &\leq \frac{\# A_{X_{n_{k+1}},X}^{(2^{k+1},2^{-(k+1)})}\cup A_{X,X_{n_{k+1}}}^{(2^{k+1},2^{-(k+1)})}}{2^{(k+1)d}}+\frac{\# A_{X_{n_k},X_{n_{k+1}}}^{(2^{k+1},2^{-(k+1)})}\cup A_{X_{n_{k+1}},X_{n_{k}}}^{(2^{k+1},2^{-(k+1)})}}{2^{(k+1)d}}\\
    &<0+\frac{1}{2^{k+1}}=\frac{1}{2^{k+1}},
    \end{split}
    \end{equation}
    where in the last step we use \eqref{eq: statcomp1} and \eqref{eq: statcomp2}. By continuing inductively, for each $l\in\N$, the estimation in \eqref{eq: statcomp3} yields the following:
    \[
    \frac{\# A_{X_{n_{k}},X}^{(2^{k+l},2^{-k})}\cup  A_{X,X_{n_{k}}}^{(2^{k+l},2^{-k})}}{2^{(k+l)d}}\leq\sum_{j=1}^{l}\frac{1}{2^{k+j}};
    \]
    This implies that:
    \begin{equation}\label{eq: statcomp4}
        \rho_{\mathsf{stat}}(X_{n_k},X)\leq 2^{-k}\sum_{j\geq 1}2^{-j}\xrightarrow[k\to\infty]{} 0.
    \end{equation}
    Therefore, from \eqref{eq: statcomp4} and since the sequence $(X_n)_{n\in\N}$ is Cauchy with respect to $\rho_{\mathsf{stat}}$, we get $\rho_{\mathsf{stat}}(X_n,X)\to 0$, as desired.
\end{proof}

\begin{rmk}
From motivations coming from dynamical systems and aperiodic order, a related (pseudo-)distance on $\mathcal{UD}_d$ is defined in \cite{autotop} by means of the formula:
\begin{equation}\label{eq: autdis}
(\forall X,Y\in\mathcal{UD}_d),\quad\rho_{\mathsf{Aut}}(X,Y):=\limsup_{L>0}\frac{\# X^{(L)}\triangle Y^{(L)}}{L^d},
\end{equation}
where $\triangle$ stands for the symmetric difference of sets; this pseudo-distance is called in \cite{autotop} the {\em autocorrelation distance}. If $\rho_{\mathsf{Aut}}(X,Y)<\epsilon<1$, the formula \eqref{eq: autdis} states that, outside of a certain large ball of radius $L_{\epsilon}>0$, the discrete sets $X,Y$ coincide statistically, meaning that the proportion of the points where they coincide in $B_L,\ L\geq L_{\epsilon}$, is $\asymp (1-\epsilon)L^d$. By considering the quotient $\mathcal{UD}_d/\sim$, where $X\sim Y$ if and only if $\rho_{\mathsf{Aut}}(X,Y)=0$, a genuine metric space is obtained.
\medskip

Observe that one could define a metric on $\mathcal{UD}_{r_0,d}$ in the same spirit as \eqref{eq: statGH} by letting $\rho_{\mathsf{Aut}}' (X,Y):=\displaystyle\sup_{L>0}\frac{\# X^{(L)}\triangle Y^{(L)}}{L^d}$. In this case, we have the relation:
\[
(\forall X,Y\in\mathcal{UD}_{r_0,d}),\quad\rho_{\mathsf{stat}}(X,Y)\leq\rho_{\mathsf{Aut}}'(X,Y).
\]

Hence, our results will be valid if we only replace the convergence with respect to $\rho_{\mathsf{stat}}$ by $\rho_{\mathsf{Aut}}'$.
\end{rmk}

Any uniformly discrete set $X$ in $\R^d$ is uniquely determined by a Radon measure by means of the embedding
\begin{equation}\label{eq: diracX}
\mathcal{UD}_d\ni X\xhookrightarrow{}\delta_X:=\sum_{p\in X}\delta_p\in\mathcal{M}(\R^d),\quad\text{where }\delta_p\text{ is the Dirac delta at }p.
\end{equation}

To warm up before addressing the stability of quasicrystallinity, we study the convergence of locally finite point configurations under both Hausdorff and statistical convergence. It is quite direct to see the following equivalence between the Hausdorff convergence of uniformly discrete sets and the weak convergence of the corresponding measures \eqref{eq: diracX}; we include its proof only to make the exposition self-contained.

\begin{prop}\label{prop: convdirac}
    Let $(X_n)_{n\geq 1}\subset\mathcal{UD}_{r_0,d}$. Then $X_n\xrightarrow{\rho_{H}}X$ if and only if $\delta_{X_n}\to\delta_X$ in the vague sense.
\end{prop}
\begin{proof}
    Let $\epsilon>0$ and fix $f\in C_c(\R^d)$ (in particular $f$ is uniformly continuous); then there is $\sigma=\sigma(\epsilon)>0$ such that:
    \[
    \|p-q\|<\sigma\quad\Rightarrow\quad\|f(p)-f(q)\|<\epsilon.
    \]
    Consider $R_f=R_{f,\epsilon}>0$ such that $\mathsf{supp}(f)\subset B_{R_f}$ and $d_H(X^{(R_f)}_n, X^{(R_f)})\leq\frac{1}{R_f}<\sigma$ for every $n\in\N$ sufficiently large; also, write $X_k^{(R_f)}=\{p_{1,n},\ldots, p_{l_{R_f},n}\}$ and $X^{(R_f)}=\{p_{1},\ldots, p_{l_{R_f}}\}$, where $\|p_{i,k}-p_i\|<\sigma$ and $l_{R_f}:=\#X^{(R_f)}=\# X_k^{(R_f)}$; this is possible since $X_n\to X$ with respect to $\rho_{H}$ and that $X_n,X\in\mathcal{UD}_{r_0,d}$, for every $n\in\N$. Then:
    \begin{equation*}
      \begin{split}
         \left|\delta_{X_n}(f)-\delta_X(f)\right|=\left|\sum_{p\in X_n}f(p)-\sum_{p\in X}f(p)\right| & =  \left|\sum_{p\in X_n^{(R_f)}}f(p)-\sum_{p\in X^{(R_f)}}f(p)\right|\\
                         & = \left|\sum_{i=1}^{l_{R_f}}f(p_{i,n})-f(p_i)\right|\\
                         & \leq\epsilon\cdot\#(\mathsf{supp}(f)\cap X),
      \end{split}
    \end{equation*}
where in the last step we used that $f$ vanishes in $B_{R_f}\setminus\mathsf{supp}(f)$; this prove the "only if" part.
\medskip

Conversely, the convergence $X_n\xrightarrow{\rho_{H}}X$ follows from the fact that the weak convergence of measures implies that $\delta_{X_n}(B_R)\to\delta_X(B_R)$, for any fixed $R>0$.
\end{proof}

To leverage Proposition \ref{prop: compvag}, we need the following lemma, which states that certain families of measures are uniformly translation-bounded.

\begin{lemma}\label{claim: unifbound}
    Let $(X_{\alpha})_{\alpha\in I}$ be a collection of elements in $\mathcal{UD}_{r_0,d}$. For each $\alpha\in I$ consider an atomic measure $\mu_{\alpha}=\sum_{p\in X_{\alpha}}\mu_{\alpha}(p)\delta_p$ supported in $X_{\alpha}$ (the $\mu_{\alpha}$ inside the sum is a function $\mu_{\alpha}:X_{\alpha}\to\C$) for which the following holds:
    \[
    \sup_{\alpha\in I}\|\mu_{\alpha}\|_{\infty}<\infty.
    \]
    Then the collection of measures $\mu_{\alpha},\ \alpha\in I$, is uniformly translation bounded.
\end{lemma}
\begin{proof}
    Let $K$ be a compact subset of $\R^d$ and $x\in\R^d$. Then, for each $\alpha\in I$, from the definition of the total variation of a measure, we have that
    \[
    |\mu_{\alpha}|(x+K)\leq\sum_{p\in X_{\alpha}}|\mu(p)|\chi_{x+
    K}(p)\leq \underbrace{\sup_{\alpha\in I}\|\mu_{\alpha}\|_{\infty}}_{=:C<\infty}\cdot\# X_{\alpha}\cap (x+K)\leq C\cdot C_{r_0,d,K},
    \]
    where $C_{r_0,d,K}$ is some positive constant depending only on $r_0, d$ and $K$, that arises from the following volume argument: since the balls $B(p,r_0/2),\ p\in X_{\alpha}$ ($\alpha$ fixed), are disjoint, we get:
    \begin{equation}\label{eq: volarg}
    \begin{split}
    \# (X_{\alpha}\cap (x+K))\mathsf{Leb}_d(B_{\frac{r_0}{2}}) & = \sum_{p\in X_{\alpha}\cap (x+K)}\mathsf{Leb}_d(B(p,r_0/2))\\
    & \leq \mathsf{Leb}_d(B(x+K,r_0/2))=\mathsf{Leb}_d(B(K,r_0/2)),
    \end{split}
    \end{equation}
    which implies that $\# (X_{\alpha}\cap (x+K))\leq \left(\frac{2}{r_0}\right)^d\frac{\mathsf{Leb}_d(B(K,r_0/2)
    )}{\mathsf{Leb}_d(B_1)}=:C_{r_0,d,K}$.
\end{proof}

The next Corollary follows directly from Proposition \ref{prop: convdirac} and \eqref{eq: obsrhostat}.

\begin{corollary}\label{cor: statvag}
    Let $X_n,\ n\geq 1$ and $X$ be uniformly discrete sets in $\mathcal{UD}_{r_0,d}$ such that $X_n\xrightarrow{\rho_{\mathsf{stat}}} X$. Then $\delta_{X_n}\to \delta_X$ vaguely.
\end{corollary}
\section{Proof of Theorem \ref{thm: A}}\label{sec: proofA}

\subsection{Fast enough convergence preserves diffractivity}\label{ssec: stabdiffdel}

Let $X$ be a uniformly discrete subset of $\R^d$. Consider the family of measures $\gamma_{X,L}$, where $L>0$, given by
\begin{equation}\label{eq: autmes}
    \gamma_{X,L}:=\frac{1}{L^d}\sum_{p,q\in X^{(L)}}\delta_{p-q}.
\end{equation}
An {\em autocorrelation measure} of $X$ is any accumulation point (in the vague sense) of the family $\{\gamma_{X,L}\}_{L>0}$. 
\medskip
The limit
\begin{equation}
    \gamma_X:=\lim_{L\to\infty}\gamma_{X,L}=\lim_{L\to\infty}\frac{1}{L^d}\sum_{p,q\in X^{(L)}}\delta_{p-q},
\end{equation}
could exist or not (compare with \eqref{eq: autoc} in the case $\mu=\delta_X$). In the positive case, we say that $X$ is {\em diffractive} and $\gamma_X$ is called the {\em autocorrelation measure} of $X$. The Fourier transform of $\gamma_X$, which is well-defined from \cite[Theorem 4.11.10]{BG2}, is called the {\em diffraction measure} of $X$, and it is denoted by $\widehat{\gamma_X}$ or $\mathcal{F}(\gamma_X)$.
\medskip

As it was discussed in the Introduction, we ask whether or not the statistical Hausdorff convergence of diffractive locally finite sets also passes, in the vague sense, to the level of the autocorrelations. In the following Proposition, we show that if a sequence of diffractive sets $X_n$ in $\mathcal{UD}_{r_0,d}$ converges rapidly enough (with respect to the statistical Hausdorff distance) to a locally finite set $X$, then $X$ is also diffractive and the sequence of the autocorrelations of $X_n$ converges to the $\gamma_X$. 

\begin{prop}\label{prop: stabdiff}
   Let $X$ be a uniformly discrete set in $\mathcal{UD}_{r_0,d}$, and $(X_n)_{n\geq 1}\subset\mathcal{UD}_{r_0,d}$ be a sequence of diffractive discrete sets in $\R^d$. Assume the following holds:
    \begin{equation}\label{eq: stabdiffconv}
        \inf\left\{\epsilon>0:\ \sup_{L>0}\frac{\#A_{X_n,X}^{(L,\epsilon)}\cup A_{X,X_n}^{(L,\epsilon)}}{L^{d/2}}<\epsilon\right\}\longrightarrow 0,\quad\text{when }n\text{ goes to }\infty.\footnote{If we consider the distance $\rho_{\mathsf{stat},1/2}$ as defined in Remark \ref{rmk statdis}, we can rewrite this condition as $X_n\xrightarrow{\rho_{\mathsf{stat},1/2}}X$.}
    \end{equation}
   Then the sequence of autocorrelation measures $(\gamma_{X_n})_{n\geq 1}$ has a unique accumulation point $\gamma$. Moreover $\gamma_X$ exists and equals $\gamma$.
\end{prop}
\begin{proof}
    Firstly, we prove that the sequence $\gamma_{X_n},\ n\geq 1$, converges (under a subsequence), to some measure $\gamma$. For this, it is sufficient to prove that $(\gamma_{X_n})_{n\geq 1}$ is uniformly translation bounded (see \eqref{eq: uniftb} and Proposition \ref{prop: compvag}); indeed, if $K$ is any compact subset of $\R^d$, then for every $x\in\R^d$:
    \begin{equation}\label{eq: stabdif1}
    \begin{split}
         |\gamma_{X_n,L}|(x+K) & =\frac{1}{L^d}\sum_{\substack{p\in X_n^{(L)}\\ -q\in X_n^{(L)}}}\chi_{x+K}(p+q)\\
                               & = \frac{1}{L^d}\sum_{-q\in X_n^{(L)}}|\delta_{X_n^{(L)}}|(x-q+K)\\
                               &\leq \frac{C_K}{L^d}\cdot|\delta_{X_n^{(L)}}|(\R^d)\\
                               & = \frac{C_K}{L^d}\cdot\# X_n^{(L)},
    \end{split}
    \end{equation}
    
    where the positive constant $C_K$ does not depend on $L$ and arises from the uniform boundedness of the measures $\delta_{X_n},\ n\geq 1$ that follows from Lemma \ref{claim: unifbound}. Now, since the $X_n$'s belong to $\mathcal{UD}_{r_0,d}$, a standard volume argument as in the proof of Lemma \ref{claim: unifbound} shows that $\# X^{(L)}_n$ is bounded from above by $L^d$ times a constant depending only on $r_0$ and $d$; specifically, we get that $\#X_n^{(L)}\leq (3L/r_0)^d$, for $L>0$ sufficiently large. 
    \medskip
    
    Hence, from the arguments of the previous paragraph and from \eqref{eq: stabdif1} we get that:
    \begin{equation*}
        |\gamma_{X_n,L}|(x+K)\leq C_K\cdot((3/r_0)^d),
    \end{equation*}

    and by letting $L\to\infty$ and taking the $\sup_n$, we obtain $\sup_{n\in\N}|\gamma_{X_n}|(x+K)$ is bounded by a constant that only depends on $K$; this proves our first claim.
    \medskip
    
    Secondly we prove that $\gamma_{X_n,L}\xrightarrow[n\to\infty]{}\gamma_{X,L}$ for any fixed $L>0$. Indeed, given $f\in C_c(\R^d)$ and $\epsilon>0$, let $\sigma\in (0,r_0)$ such that $|f(p)-f(q)|<\epsilon$ provided $\|p-q\|<\sigma$. By \eqref{eq: stabdiffconv}, there exists $n_{\epsilon}\in\N$ such that, for every $n\geq n_{\epsilon}$ it holds that

    \begin{equation}\label{eq: stabdiff0}
        (\forall L>0),\quad\# A_{X_n,X}^{(L,\sigma/2)}\cup A_{X,X_n}^{(L,\sigma/2)}<\frac{\sigma}{2}L^{d/2}
    \end{equation}
    Then, since $X_n, X\in\mathcal{UD}_{r_0,d}$, we can write
    \[
    X_n^{(L)}\setminus A_{X_n,X}^{(L,\sigma/2)}=\{p_{1,n},\ldots, p_{k_L,n}\}\quad\text{and}\quad X^{(L)}\setminus A_{X,X_n}^{(L,\sigma/2)}=\{p_1,\ldots,p_{k_L}\},
    \]
    for some $k_L>0$ only depending on $L, r_0$ and the dimension $d$. Moreover, for each $i\in\{1,\ldots, k_L\}$, write
    \[
    X_n^{(L)}\cap(\mathsf{supp}(f)+p_{i,n})=\{p_{1,n}^{i},\ldots, p_{l_i,n}^{i}\}\quad\text{and}\quad X^{(L)}\cap(\mathsf{supp}(f)+p_{i})=\{p_{1}^{i},\ldots, p_{l_i}^{i}\},
    \]
    where $l_i\leq\min\{k_L, \underbrace{\#\mathsf{supp}(f)\cap X^{(L)}}_{=:k_f}\}$. Thus, we obtain that: 
    \begin{equation}\label{eq: stabproof}
        \begin{split}
            \left|\gamma_{X_n,L}(f)-\gamma_{X,L}(f)\right| & = \frac{1}{L^d}\left|\sum_{p,q\in X_n^{(L)}}f(p-q)-\sum_{p,q\in X^{(L)}}f(p-q)\right|\\
            &\leq\frac{1}{L^d}\left|\sum_{i=1}^{k_L}\sum_{j=1}^{l_i}f(p_{j,n}^{i}-p_{i,n})-\sum_{i=1}^{k_L}\sum_{j=1}^{l_i}f(p_j^{i}-p_i)\right|\\
            &+\sum_{p,q\in A_{X_n,X}^{(L,\sigma/2)}\cup A_{X,X_n}^{(L,\sigma/2)}}\left|f(p-q)\right|\\
            &\leq\frac{1}{L^d}\sum_{i=1}^{k_L}\sum_{j=1}^{l_i}|f(p_{j,n}^{i}-p_{i,n})-f(p_j^{i}-p_i)|+\frac{\|f\|_{\infty}}{L^d}\cdot \#\left(A_{X_n,X}^{(L,\sigma/2)}\cup A_{X,X_n}^{(L,\sigma/2)}\right)^2\\
            & \leq\frac{\epsilon\cdot k_Lk_f}{L^d}+\|f\|_{\infty}\frac{\sigma^2}{4},
        \end{split}
    \end{equation}
 where in the last step we used \eqref{eq: stabdiffconv} for $n\geq n_{\epsilon}$ sufficiently large, $l_i\leq k_f$, and that
  \[
  \|(p_{i,n}-p_{j,n}^{i})-(p_i-p_j^{i})\|\leq \|p_{i,n}-p_i\|+\|p_{j,n}^{i}-p_j^{i}\||<\frac{\sigma}{2}+\frac{\sigma}{2}=\sigma.
  \]
  Hence $\gamma_{X_n,L}\xrightarrow[n\to\infty]{}\gamma_{X,L}$ in the vague sense.\footnote{Actually, the convergence $\gamma_{X_n,L}\to\gamma_{X,L}$ works if $X_n\xrightarrow{\rho_{\mathsf{stat}}} X$, but the upper bound at the end of \eqref{eq: stabproof} will depend on multiplication by $L$.}
  \medskip
  
  Now, let $\gamma$ be a vague accumulation point of the sequence $\gamma_{X_n},\ n\geq 1$, and $\gamma_{X_{n_k}},\ k\in\N$ such that $\gamma_{X_{n_k}}\to\gamma$. To show that $\gamma_{X,L}$ converges vaguely to $\gamma$, let $\epsilon>0$ be given and $f\in C_c(\R^d)$. Let $n_{\epsilon}\in\N$ be sufficiently large such that \eqref{eq: stabproof} and $|\gamma_{X_{n_k}}(f)-\gamma(f)|<\epsilon$ hold for $n_k\geq n_{\epsilon}$; it should be noticed at this point that the choice of $n_{\epsilon}$ does not depend on $L$, but only on $f$ and $\sigma$, which in turn depends on $\epsilon$. Moreover, let $L>0$ large-enough such that $|\gamma_{X_{n_k},L}(f)-\gamma_{X_{n_k}}(f)|<\epsilon$ and $\mathsf{supp}(f)\subset B_L$ ($L$ depends on $n_{\epsilon}$ too). Then, by triangle inequality, we get:
  \begin{equation*}
     \begin{split}
       |\gamma_{X,L}(f)-\gamma(f)| & \leq |\gamma_{X,L}(f)-\gamma_{X_{n_k},L}(f)|+|\gamma_{X_{n_k},L}(f)-\gamma_{X_{n_k}}(f)|+|\gamma_{X_{n_k}}(f)-\gamma(f)|\\
       &<\left(\frac{\epsilon k_Lk_f}{L^d}+\|f\|_{\infty}\frac{\sigma^2}{4}\right)+2\epsilon\\
       &\leq \frac{\# X^{(L)}}{L^d}\cdot\epsilon k_f+\|f\|_{\infty}\frac{\sigma^2}{4}+2\epsilon\leq \epsilon \cdot C_{r_0,d} k_f+\frac{\sigma^2}{4}+2\epsilon.
     \end{split}
  \end{equation*}
  
  This finishes the proof since $\sigma\to 0$ when $\epsilon\to 0$.
\end{proof}

\begin{rmk}\label{rmk: notcongh}
\begin{enumerate}
    \item Proposition \ref{prop: stabdiff} is no longer true if we replace \eqref{eq: stabdiffconv} by $\rho_{H}$; indeed, just consider the sequence $(X_n)_{\geq 1}$ given by 
    \[
    X_n^{(n)}=\Z^2\cap B_n\quad\text{and}\quad X_n\setminus B_n=\mathsf{Pen}\setminus B_n,
    \]
    where $\mathsf{{Pen}}\subset\R^2$ is the Penrose lattice obtained by putting one point in the center of every tile of the Penrose tiling \cite{penrose}. Observe that $X_n\xrightarrow{\rho_{H}}\Z^2$, and in addition 
    \[
    (\forall n\in\N),\quad\gamma_{X_n}=\gamma_{\mathsf{Pen}},
    \]
    since $X_n$ and $\mathsf{Pen}$ only differ in a bounded region. However, we have that $\gamma_{X_n}=\gamma_{\mathsf{Pen}}\to\gamma_{\mathsf{Pen}}\neq\gamma_{\Z^2}$.
    
    \item It would be interesting to know whether or not the convergence rate in \eqref{eq: stabdiffconv} can be improved. In other words, we ask if there is $\alpha\in\left(0,\frac{1}{2}\right)$ such that the assumption \eqref{eq: stabdiffconv} can be replaced by 
    \[
    \rho_{\mathsf{stat},\alpha}(X_n,X)\to 0.
    \]
\end{enumerate}
     
\end{rmk}
\subsection{On spectral stability under convergence}\label{ssec: specstab}
In this Section, we focus on whether the spectral type (of the diffraction measure) of quasicrystals is stable under vague convergence of autocorrelations. It is known that the spectral type of a measure can vary under vague convergence; see, for instance, \cite[Example 8.8]{aper}. A uniformly discrete set $X$ that is diffractive is said to be a {\em mathematical quasicrystal} if 
\begin{equation}\label{eq: quasicrys}
    \widehat{\gamma_X}\perp\mathsf{Leb}_d.
\end{equation}

We say that a family $(X_{\alpha})_{\alpha\in I}$ of diffractive subsets in $\mathcal{UD}_{r_0,d}$ is {\em uniformly diffractive} if the there exists a measure $\mu$ such that 

\begin{equation}\label{eq: unifquasicrys}
    (\forall\alpha\in I),\quad\widehat{\gamma_{X_{\alpha}}}\ll\mu.
\end{equation}

A sequence of uniformly diffractive point sets can be easily constructed: from a diffractive set $X$ in $\R^d$, for each $n\in\N$, take an arbitrary locally finite set $X_n'\subset\R^d$ disjoint from $X$ and such that 
\[
\lim_{L\to\infty}\frac{\# X_n'\cap B_L}{L^d}=0,
\]
and then define $X_n:=X\cup X_n'$, while keeping the uniform discreteness of the $X_n$'s; each element of this sequence of locally finite point sets has the same autocorrelation as $X$, and so the same diffraction measure.
\medskip

Let $\gamma,\gamma'$ be two Radon measures in $\R^d$. We define the {\em localized Hellinger density} (or the {\em localized affinity}) $h(\gamma,\gamma')$ in duality with compactly supported functions as follows:
\begin{equation}\label{eq: helldens}
 (\forall f\in C_c(\R^d)),\quad\langle h(\gamma,\gamma'),f\rangle =\int_{\R^d}\left(\frac{d\gamma}{d\lambda}(x)\right)^{1/2}\left(\frac{d\gamma'}{d\lambda}(x)\right)^{1/2}f(x)d\lambda(x),    
\end{equation}

where the integral is taken with respect to a Radon measure $\lambda$ such that $\gamma, \gamma'\ll\lambda$; this measure always exists, since $\gamma\ll\gamma+\gamma'$ and $\gamma'\ll\gamma+\gamma'$. Moreover, the Hellinger density is independent of the choice of the measure $\lambda$. We use the following fact about the Hellinger density: $\gamma\perp\gamma'$ if and only if $h(\gamma,\gamma')=0$; see \cite{mespec} and \cite[A.2]{almostsure1} for further details.
\medskip

With this setting in mind, we are in a position to state our following result regarding the stability of the quasi-crystalline nature of uniformly diffractive sequences of discrete sets under convergence.
    
\begin{prop}\label{prop: stabspec}
    Let $X_n,\ n\in\N$, and $X$ be diffractive subsets of $\R^d$ such that $\gamma_{X_n}\to\gamma_X$ vaguely. Then $\widehat{\gamma_{X_n}}\to\widehat{\gamma_X}$ vaguely too.  Moreover, if the sequence $(X_n)_{n\geq 1}$ is uniformly diffractive satisfying that
     \[
     h(\widehat{\gamma_{X_n}},\mathsf{Leb}_d)\to 0\quad\text{vaguely},
     \] then $X$ is a quasicrystal.
\end{prop}

\begin{proof}
    The first part of the Proposition follows from \cite[Theorem 4.11.10]{BG2}, since the $\gamma_{X_n}$'s are positive definite Fourier transformable measures, so $\gamma_X$ it is too, and
    \[
    \mathcal{F}(\gamma_X)=\mathcal{F}(\lim_{n\to\infty}\gamma_{X_n})=\lim_{n\to\infty}\mathcal{F}(\gamma_{X_n}).
    \]
    
    For the second part of the Proposition, we need to prove that $\widehat{\gamma_X}\perp\mathcal{L}$, where we use the notation $\mathcal{L}=\mathsf{Leb}_d$. Since $(X_n)_{n\geq 1}$ is uniformly quasicrystalline, we can choose a Radon measure $\lambda$ such that, for every $n\in\N$:
    \[
    \widehat{\gamma_{X_n}}\ll\lambda,\quad \widehat{\gamma_X}\ll\lambda\quad\text{and}\quad\mathcal{L}\ll\lambda;
    \]
    
the above is possible just by taking, for instance, $\lambda=\mu+\widehat{\gamma_X}+\mathcal{L}$, where $\mu$ is the measure in \eqref{eq: unifquasicrys}, and by considering the discussion in the paragraph below \eqref{eq: helldens}. Let us denote the Radon-Nykodim derivatives of $\widehat{\gamma_{X_n}}$ and $\widehat{\gamma_{X}}$ with respect to $\lambda$ by $\frac{d\widehat{\gamma_{X_n}}}{d\lambda}$ and $\frac{d\widehat{\gamma_X}}{d\lambda}$, respectively. First we shall prove, for every $f\in C_c(\R^d)$, that $\frac{d\widehat{\gamma_{X_n}}}{d\lambda}\cdot f$ converges in $L^1(\lambda)$ to $\frac{d\widehat{\gamma_X}}{d\lambda}\cdot f$. This follows since the $\widehat{\gamma_{X_n}}$'s are absolutely continuous with respect to $\lambda$, and because the vague convergence $\widehat{\gamma_{X_n}}\to\widehat{\gamma_X}$; indeed:
    \begin{equation*}
        \begin{split}
             \int_{\R^d}\frac{d\widehat{\gamma_{X_n}}}{d\lambda}(x)f(x)d\lambda(x)=\widehat{\gamma_{X_n}}(f)\to\widehat{\gamma_X}(f)=\int_{\R^d}\frac{d\widehat{\gamma_{X}}}{d\lambda}(x)f(x)d\lambda(x).
        \end{split}
    \end{equation*}

 From this, we now prove that $\langle h(\widehat{\gamma_{X_n}},\mathcal{L}),f\rangle\to\langle h(\widehat{\gamma_X},\mathcal{L}),f\rangle$\footnote{In general, if we have $\gamma_n\to\gamma$ and $\gamma_n'\to\gamma'$, then we can ensure only that $\displaystyle\limsup_{n}\langle h(\gamma_n,\gamma_n'),f\rangle\leq\langle h(\gamma,\gamma'),f\rangle$; see \cite[Lemma A.5]{almostsure1} and \cite[Theorem B]{almostsure3}} for every compactly supported continuous function $f$. This is equivalent to showing that
    \begin{equation}\label{eq: stabspec1}
        \left|\int_{\R^d}\left(\frac{d\widehat{\gamma_{X_n}}}{d\lambda}(x)\right)^{1/2}\left(\frac{d\mathcal{L}}{d\lambda}(x)\right)^{1/2}f(x)d\lambda(x)-\int_{\R^d}\left(\frac{d\widehat{\gamma_{X}}}{d\lambda}(x)\right)^{1/2}\left(\frac{d\mathcal{L}}{d\lambda}(x)\right)^{1/2}f(x)d\lambda(x)\right|\to 0
    \end{equation}

     By the inequality $|\sqrt{a}-\sqrt{b}|\leq\sqrt{|a-b|},\ a,b\geq 0$, and the Cauchy-Schwarz inequality, the left hand side of \eqref{eq: stabspec1} can be bounded from above by:
    \begin{equation}\label{eq: specstab2}
        \begin{split}
            &\int_{\R^d}\left(\frac{d\mathcal{L}}{d\lambda}(x)|f(x)|\right)^{1/2}\left|\left(\frac{d\widehat{\gamma_{X_n}}}{d\lambda}(x)|f(x)|\right)^{1/2}-\left(\frac{d\widehat{\gamma_{X}}}{d\lambda}(x)|f(x)|\right)^{1/2}\right|d\lambda(x)\\
            &\leq\int_{\R^d}\left(\frac{d\mathcal{L}}{d\lambda}(x)|f(x)|\right)^{1/2}\left|\left(\frac{d\widehat{\gamma_{X_n}}}{d\lambda}(x)\right)|f(x)|-\left(\frac{d\widehat{\gamma_{X}}}{d\lambda}(x)\right)|f(x)|\right|^{1/2}d\lambda(x)\\
            &\leq\left(\int_{\R^d}\frac{d\mathcal{L}}{d\lambda}(x)|f(x)|d\lambda(x)\right)^{1/2}\left(\int_{\R^d}\left|\left(\frac{d\widehat{\gamma_{X_n}}}{d\lambda}(x)\right)|f(x)|-\left(\frac{d\widehat{\gamma_{X}}}{d\lambda}(x)\right)|f(x)|\right|d\lambda(x)\right)^{1/2}.
        \end{split}
    \end{equation}

    Since $\frac{d\widehat{\gamma_{X_n}}}{d\lambda}|f|\xrightarrow{L^1(\lambda)}\frac{d\widehat{\gamma_{X}}}{d\lambda}|f|$, then the last bound in \eqref{eq: specstab2} converges to $0$; this implies the convergence 
    \[
    (\forall f\in C_c(\R^d)),\quad\langle h(\widehat{\gamma_{X_n}},\mathcal{L}),f\rangle\to\langle h(\widehat{\gamma_X},\mathcal{L}),f\rangle.
    \]
    Finally, since $\langle h(\widehat{\gamma_{X_n}},\mathcal{L}),f\rangle\to 0$, the equality $h(\widehat{\gamma_X},\mathcal{L})=0$ follows; in other words, $\widehat{\gamma_X}\perp\mathcal{L}$ as claimed.
\end{proof}

\begin{rmk}
    In the proof of Proposition \ref{prop: stabspec}, the spectral type of $(\widehat{\gamma_{X_n}})_{n\in\N}$ may change. Indeed, it could happen that the diffraction measure $\widehat{\gamma_X}$ has a singular continuous spectrum, while the $\widehat{\gamma_{X_n}}$'s have a purely point one; the important thing about the result is that still we have $\widehat{\gamma_X}\perp\mathsf{Leb}_d$. 
\end{rmk}
\begin{proof}[Proof of Theorem \ref{thm: A}]
    It follows directly by putting together Propositions \ref{prop: stabdiff} and \ref{prop: stabspec}. 
\end{proof}

\section{Proof of Theorem \ref{thm: B}}\label{sec: proofB}

Recall that, given a locally finite set $X\subset\R^d$, its {\em Fourier Transform} $\mathcal{F}_X$ is defined by the following limit in the vague sense: for every $f\in C_c(\R^d)$
\begin{equation}\label{eq: FXdef}
\mathcal{F}_X(f):=\lim_{L\to\infty}\frac{1}{L^d}\widehat{\delta_{X^{(L)}}}(f)=\lim_{L\to\infty}\frac{1}{L^d}\sum_{p\in X^{(L)}}\int_{\R^d}e^{-2\pi i\langle p,x\rangle}f(x)dx.
\end{equation}

Similarly to the discussion of the existence of the limit defining \eqref{eq: autmes}, we cannot always ensure that the limit in \eqref{eq: FXdef} does exist, but we can prove that the family $\widehat{\delta_{X^{(L)}}}/L^d,\ L>0$, is precompact.

\begin{lemma}\label{lem: FXmeas}
For any $X\in\mathcal{U}_{r_0,d}$ the family $(\mathcal{F}_{X,L})_{L>0}$ given by 
\begin{equation}\label{eq: FXmeas}
\mathcal{F}_{X,L}:=\frac{1}{L^d}\widehat{\delta_{X^{(L)}}},
\end{equation}
is precompact in $\mathcal{M}(\R^d)$.
\end{lemma}

\begin{proof}
Given $X\in\mathcal{U}_{r_0,d}$, we shall prove that the family $\mathcal{F}_{X,L},\ L>0$ is vaguely bounded; then the conclusion follows by Proposition \ref{prop: compvag}.
\medskip

Let us consider $f\in C_c(\R^d)$, and let $C_{r_0,d}$ be a non-negative constant, depending only on $r_0,d$, such that 
\[
\frac{\# X^{(L)}}{L^d}\leq C_{r_0,d},
\]
which can be found from the uniform discreteness of $X$, following the very same volume argument as in \eqref{eq: volarg}. Then, we have that
\begin{equation}\label{eq: FXmeasunif}
\begin{split}
\left|\frac{1}{L^d}\widehat{\delta_{X^{(L)}}}(f)\right| & = \frac{1}{L^d}\left|\delta_{X^{(L)}}(\widehat{f})\right|\\
                                        & =\frac{1}{L^d}\left|\sum_{p\in X^{(L)}}\int_{\R^d}e^{-2\pi i\langle p,x\rangle}f(x)dx\right|\\
                                        & \leq\frac{\|f\|_{L^1(dx)}}{L^d}\cdot \# X^{(L)}\leq C_{r_0,d}\|f\|_{L^1(dx)}.
\end{split}
\end{equation}

Thus, we get $\displaystyle\sup_{L>0}\mathcal{F}_{X,L}(f)<\infty$, and thus $\left(\mathcal{F}_{X,L}\right)_{L>0}$ is vaguely bounded, as claimed. 
\end{proof}
Henceforth, let us denote by $\mathcal{Q}_{r_0,d}$ the subset of all the uniform discrete point sets $X\in\mathcal{UD}_{r_0,d}$ such that the limit \eqref{eq: FXdef} defining $\mathcal{F}_X$ exists.

\begin{lemma}\label{cor: Mprecom}
Let $\mathcal{M}_{r_0,d}$ be the set of the Fourier Transforms of the $X$'s belonging $\mathcal{Q}_{r_0,d}$. Then $\mathcal{M}_{r_0,d}$ is precompact.
\end{lemma}

\begin{proof}
The proof follows from \eqref{eq: FXmeasunif}, since we can use the very same constant $C_{r_0,d}\geq 0$ as in Lemma \ref{lem: FXmeas} to get, for any $f\in C_c(\R^d)$, that:
\begin{equation*}
\frac{1}{L^d}\left|\widehat{\delta_{X^{(L)}}}(f)\right|\leq C_{r_0,d}\|f\|_{L^1(dx)},
\end{equation*}
which implies that
\begin{equation}\label{eq: Mprecom}
\left|\mathcal{F}_X(f)\right|\leq C_{r_0,d}\|f\|_{L^1(dx)}.
\end{equation}

Since the upper bound in \eqref{eq: Mprecom} does not depend on the precise geometry of $X$, but only on the separation radius $r_0$ and the dimension $d$, then it holds that:
\[
(\forall f\in C_c(\R^d)),\quad\sup_{X\in\mathcal{Q}_{r_0,d}}\left|\mathcal{F}_X(f)\right|<\infty.
\]

Then again the conclusion follows from Proposition \ref{prop: compvag}.
\end{proof}
Firstly, we show that $\mathcal{Q}_{r_0,d}$ is closed.

\begin{prop}\label{prop: Qclosed}
$\mathcal{Q}_{r_0,d}$ is a closed subset of $(\mathcal{UD}_{r_0,d},\rho_{\mathsf{stat}})$.
\end{prop}

\begin{proof}
Let $(X_n)_{n\geq 1}$ be a sequence in $\mathcal{Q}_{r_0,d}$ such that $X_n\xrightarrow{\rho_{\mathsf{stat}}} X\in\mathcal{UD}_{r_0,d}$; we shall prove that $X\in\mathcal{Q}_{r_0,d}$. By Lemma \ref{cor: Mprecom}, the sequence $(\mathcal{F}_{X_n})_{n\geq 1}$ possesses at least one convergent subsequence, that we denote by $(\mathcal{F}_{n_k})_{k\geq 1}$ and its limit by $\mathcal{L}$. We now verify that $\mathcal{F}_X$ exists and
\begin{equation}\label{eq: Qclosed}
\mathcal{L}=\lim_{k\to\infty}\mathcal{F}_{X_{n_k}}=\mathcal{F}_X\quad\text{vaguely}.
\end{equation}

For this, let us fix $f\in C_c(\R^d)$ and $\epsilon>0$. Then, by triangle inequality we have the following for every $k\in\N$:
\begin{equation}\label{eq: QclosedE}
\begin{split}
\left|\frac{1}{L^d}\widehat{\delta_{X^{(L)}}}(f)-\mathcal{L}(f)\right| & \leq \underbrace{\frac{1}{L^d}\left|\widehat{\delta_{X_{n_k}^{(L)}}}(f)-\widehat{\delta_{X^{(L)}}}(f)\right|}_{=(E1)}+\underbrace{\left|\frac{1}{L^d}\widehat{\delta_{X_{n_k}^{(L)}}}(f)-\mathcal{L}(f)\right|}_{=(E2)}.
\end{split}
\end{equation}

To estimate $(E1)$ in \eqref{eq: QclosedE}, let $R_f>0$ be the radius of the smallest closed ball centered at the origin containing $\mathsf{supp}(f)$; let also $\epsilon\in\left(0,2\pi R_fr_0\right)$. For $x\in\mathsf{supp}(f)$, from the inequality $|\sin t-\sin s|\leq |t-s|$, the following holds:
    \begin{equation*}
    \begin{split}
    |e^{-2\pi i\langle p,x\rangle}-e^{-2\pi i\langle q,x\rangle}| & \leq |\cos (2\pi\langle p,x\rangle)-\cos(2\pi \langle q,x\rangle)|+|\sin (2\pi\langle p,x\rangle)-\sin(2\pi \langle q,x\rangle)|\\
    &\leq 4\pi|\langle p-q,x\rangle|\\
    &\leq 4\pi R_f\|p-q\|,
    \end{split}
    \end{equation*}
    Thus, if we take $\sigma=\epsilon/4\pi R_f<r_0/2$, we get the following for every $x\in\R^d$: 
    \begin{equation}\label{eq: unifexp}
        \|p-q\|<\sigma\quad\Rightarrow\quad |e^{-2\pi i\langle p,x\rangle}-e^{-2\pi i\langle q,x\rangle}|<\epsilon.
    \end{equation}
    
    For this choice of $\sigma$ and by the convergence $X_n\xrightarrow{\rho_{\mathsf{stat}}}X$, we have there exists $n_{\epsilon}\in\N$ such that:
    \begin{equation}\label{eq: conFT0}
    (\forall n\geq n_{{\epsilon}}),\quad\sup_{L>0}\frac{\# A_{X_n,X}^{(L,\sigma)}\cup A_{X,X_n}^{(L,\sigma)}}{L^d}<\sigma.
    \end{equation}
    Then, by the boundedness of $e^{-2\pi i\langle\cdot,x\rangle}$, the triangle inequality and \eqref{eq: conFT0} we have the following estimations:
    
    \begin{equation}\label{eq: convFT}
    \begin{split}
          \frac{1}{L^d}\left|\widehat{\delta_{X_{n_k}^{(L)}}}(f)-\widehat{\delta_{X^{(L)}}}(f)\right| & = \frac{1}{L^d}\left|\sum_{p\in X_{n_k}^{(L)}}\int_{\R^d} e^{-2\pi i\langle p,x\rangle}f(x)dx-\sum_{p\in X^{(L)}}\int_{\R^d} e^{-2\pi i\langle p,x\rangle}f(x)dx\right|\\
          & \leq\frac{1}{L^d}\int_{\mathsf{supp}(f)}\left|\sum_{p\in X_{n_k}^{(L)}}e^{-2\pi i\langle p,x\rangle}-\sum_{p\in X^{(L)}} e^{-2\pi i\langle p,x\rangle}\right||f(x)|dx\\
          & \leq\frac{1}{L^d}\int_{\mathsf{supp}(f)}\left|\sum_{p\in X_{n_k}^{(L)}\setminus A_{X_{n_k},X}^{(L,\sigma)}}e^{-2\pi i\langle p,x\rangle}-\sum_{p\in X^{(L)}\setminus A_{X,X_{n_k}}^{(L,\sigma)}} e^{-2\pi i\langle p,x\rangle}\right||f(x)|dx\\
          & + 2\left(\frac{\# A_{X_{n_k},X}^{(L,\sigma)}\cup A_{X,X_{n_k}}^{(L,\sigma)}}{L^d}\right)\|f\|_{L^1(dx)}\\
          &\leq \frac{1}{L^d}\int_{\mathsf{supp}(f)}\left(\sum_{j=1}^{N_L}|e^{-2\pi i\langle p_{j,{n_k}},x\rangle}-e^{-2\pi i\langle p_j,x\rangle}|\right) |f(x)|dx+2\sigma\|f\|_{L^1(dx)}\\
          &\leq \|f\|_{L^{1}(dx)}\left(\frac{\epsilon N_L}{L^d}+2\sigma\right)<\|f\|_{L^1(dx)} C_{r_0,d}(\epsilon+2\sigma),
          \end{split}
\end{equation}

    where $N_L$ is bounded from above by $C_{r_0,d}\cdot L^d$, and $C_{r_0,d}$ is a positive constant depending only on $r_0$ and $d$ (actually, we have that $N_L\leq \# X^{(L)}$), and the $p_{j,{n_k}}$'s and $p_j$'s are given in \eqref{eq: sigmaclose}. Therefore we obtain:
\begin{equation}\label{eq: QclosedE1}
    (\forall n_k\geq n_{\epsilon}),\quad\frac{1}{L^d}\left|\widehat{\delta_{X_{n_k}^{(L)}}}(f)-\widehat{\delta_{X^{(L)}}}(f)\right|<C_{r_0,d}\|f\|_{L^{1}(dx)}(\epsilon+2\sigma),
\end{equation}
    
This yields $(E1)\xrightarrow[\epsilon\to 0] {}0$ by noting that $\sigma(\epsilon)\xrightarrow[\epsilon\to 0]{}0$.\\ 

To estimate $(E2)$ in \eqref{eq: QclosedE} for $\epsilon>0$ as above, by the convergence $\mathcal{F}_{X_{n_k}}\to\mathcal{L}$, there exists a positive integer $k_{\epsilon}$ such that $n_{k_{\epsilon}}\geq n_{\epsilon}$ large-enough such that 
\begin{equation}\label{eq: QclosedE21}
(\forall k\geq k_{\epsilon}),\quad\left|\mathcal{F}_{X_{n_k}}(f)-\mathcal{F}(f)\right|<\epsilon.
\end{equation}
On the other hand, since $\frac{1}{L^d}\widehat{\delta_{X_{n_k}^{(L)}}}\to\mathcal{F}_{X_{n_k}}$ when $L$ goes to $\infty$, there is $L_{\epsilon}=L(k_{\epsilon})$ such that
\begin{equation}\label{eq: QclosedE22}
\left|\frac{1}{L^d}\widehat{\delta_{X_{n_{k_{\epsilon}}}^{(L)}}}(f)-\mathcal{F}_{X_{n_{k_{\epsilon}}}}(f)\right|<\epsilon.
\end{equation}

Thus, by triangle inequality together \eqref{eq: QclosedE21} and \eqref{eq: QclosedE22}, we obtain that 
\begin{equation}\label{eq: QclosedE2}
\left|\frac{1}{L^d}\widehat{\delta_{X_{n_{k_\epsilon}}^{(L)}}}(f)-\mathcal{L}(f)\right|<2\epsilon.
\end{equation}

We finish the proof by putting together \eqref{eq: QclosedE1} and \eqref{eq: QclosedE2}, with the fact their regimes are fully compatible since the upper bound in \eqref{eq: QclosedE1} does not depend on $L$.
\end{proof}

Now we are in conditions of addressing the continuity of the operator $\mathcal{F}:\mathcal{Q}_{r_0,d}\ni X\mapsto\mathcal{F}_X\in\mathcal{M}(\R^d)$, where $\mathcal{Q}_{r_0,d}$ is endowed with the distance $\rho_{\mathsf{stat}}$ (restricted to $\mathcal{Q}_{r_0,d}$) and $\mathcal{M}(\R^d)$ with the vague topology. 
\begin{prop}\label{lemma: convFT}
    Let $X_n$, with $n\in\N$, and $X$ be uniformly discrete sets in $\mathcal{Q}_{r_0,d}$. If $X_n\xrightarrow{\rho_{\mathsf{stat}}}X$, then:
    \[
    \quad\mathcal{F}_{X_n}\to\mathcal{F}_X\quad\text{vaguely}.
    \] 
\end{prop}

\begin{proof}
For $X_n\xrightarrow{\rho_{\mathsf{stat}}} X$ as in the statement of the Proposition, we already know that $X$ must belong to $\mathcal{Q}_{r_0,d}$, and thus $\mathcal{F}_X$ necessarily exists. It only remains to show that
\[
\mathcal{F}_{X_n}\to\mathcal{F}_X\quad\text{vaguely}.
\]
However this essentially follows the very same computations as in \eqref{eq: convFT}, and this concludes the Proposition.
\end{proof}
\begin{rmk}
    A similar observation as in \ref{rmk: notcongh} shows that Proposition \ref{lemma: convFT} is no longer true if we consider $\rho_{H}$ instead $\rho_{\mathsf{stat}}$.
\end{rmk}

\section*{Acknowledgement}

I would like to thank Mircea Petrache for comments on the Introduction to this article.

\appendix 

\section{Fourier recoveraibility: proof of Theorem \ref{thm: C}}\label{ssec: stabrob}

Now we address the problem of stability under perturbations. Recall that, given a uniformly discrete set $X\in\mathcal{UD}_d$, for each $p\in X$ we consider a random vector $\xi_p:\Omega\to\R^d$, such that the $\xi_p$'s are iid modelled in the same probability space $(\Omega,\mathcal{T},\mathbb{P})$ with common law $\xi$. Then we consider the {\em random perturbation} of $X$ given by
\[
X_{\xi}:=\{p+\xi_p:\ p\in X\}.
\]

We say that a locally finite set $X\subset\R^d$ has an {\em asymptotic density} $\mathsf{dens}>0$ if the following limit exists:
\begin{equation}\label{eq: asdens}
    \lim_{L\to\infty}\frac{\# X^{(L)}}{L^d}=:\mathsf{dens}.
\end{equation}

Given a sequence $(X_n)_{n\geq 1}\subset\mathcal{UD}_{r_0,d}$,  $X\in\mathcal{UD}_{r_0,d}$ and $0<\sigma<r_0/2$., we adopt the following notation:
\begin{equation}\label{eq: sigmaclose}
    X_n\setminus A_{X_n,X}^{(L,\sigma)}=\{p_{1,n},\ldots, p_{k_L,n}\}\quad\text{and}\quad X\setminus A_{X,X_n}^{(L,\sigma)}=\{p_1,\ldots,p_{k_L}\};
\end{equation}
in particular, from \eqref{eq: sigmaclose} it holds that $\|p_{i,n}-p_i\|<\sigma$, for every $i\in\{1,\ldots, k_L\}$.
\medskip

We require the following generalization of \cite[Lemma 3.1]{almostsure3} (see also \cite{mespec}).
 
\begin{lemma}\label{lem: controlperb}
Let $(X_n)_{n\in\N}\in\mathcal{UD}_{r_0,d}$ be a sequence of uniformly discrete sets, such that $X_n$ has asymptotic density for every $n\in\N$. Suppose that there exists a positive number $L_0$ such that 
\begin{equation}\label{eq: uniffar}
    (\forall n\in\N),\quad X_n\setminus\{0\}\cap B_{L_0}=\emptyset.
\end{equation}
For each $n\in\N$, let $\xi_p^{(n)},\ p\in X_n$ be iid random vectors, with $\xi_p^{(n)}\sim\xi^{(n)}$ and such that 
\begin{equation}\label{eq: unifnoise}
    \sup_{n\in\N}\mathbb{E}(\|\xi^{(n)}\|^{d+\epsilon})<\infty\quad\text{for some }\epsilon>0.
\end{equation}
Then, almost surely, the following limits hold:
\begin{equation}\label{eq: controlperb}
    \lim_{L\to\infty}\sup_{n\in\N}\frac{\#\{p\in X_n^{(L)}:\ \|p+\xi_p^{(n)}\|>L\}}{L^d}=0\quad\text{and}\quad\lim_{L\to\infty}\sup_{n\in\N}\frac{\#\{p\in X_n\setminus B_L:\ \|p+\xi_p^{(n)}\|< L\}}{L^d}=0.
\end{equation}
\end{lemma}
\begin{proof}
      Let $\delta:=\frac{\epsilon}{2(d+\epsilon)}$. Then, by Chevyshev's inequality, we have the following for every non-zero vector $p\in X_n$:
      \begin{equation*}
          \begin{split}
              \mathbb{P}(\|\xi_p^{(n)}\|\geq \|p\|^{1-\delta})\leq\frac{\mathbb{E}(\|\xi_p^{(n)}\|^{d+\epsilon})}{\|p\|^{(d+\epsilon)(1-\delta)}}=\frac{\mathbb{E}(\|\xi_p^{(n)}\|^{d+\epsilon})}{\|p\|^{d+\frac{\epsilon}{2}}}.
          \end{split}
      \end{equation*}
      Therefore, we have that
      \begin{equation}\label{eq: perturproof}
      \sum_{p\in X_n}\mathbb{P}(\|\xi_p^{(n)}\|\geq\|p\|^{1-\delta})\leq 1+\mathbb{E}(\|\xi^{(n)}\|^{d+\epsilon})\sum_{p\in X_n\setminus\{0\}}\|p\|^{-d-\frac{\epsilon}{2}}\leq C_1<\infty,
       \end{equation}
      where $C_1$ is a positive constant that does not depend on $n\in\N$, which is obtained from \eqref{eq: unifnoise} and the uniform bound for $\sum_{p\in X_n\setminus\{0\}}\|p\|^{-d-\frac{\epsilon}{2}},\ n\in\
      N$, given from \eqref{eq: uniffar} and the fact that the $X_n$'s belong to $\mathcal{UD}_{r_0,d}$. 
      \medskip

      Let us define $F_n\geq 0$ by
      \[
      F_n:=\#\{p\in X_n:\ \|\xi_p^{(n)}\|\geq \|p\|^{1-\delta}\};
      \]
      then, from \eqref{eq: perturproof}, the Borel-Cantelli Lemma tells us that almost surely the following holds:
      \[
      \sup_{n\in\N} F_n<\infty. 
      \]
      Thus, by the triangle inequality, we get that
      \[
      \#\{p\in X_n^{(L)}:\ \|p+\xi_p^{(n)}\|>L\}\leq \#\{p\in X_n^{(L)}:\ \|p+\xi_p^{(n)}\|>L,\ \|\xi_p^{(n)}\|<\|p\|^{1-\delta}\}+F_n;
      \]
      On the other hand,
      \[
      \#\{p\in X_n^{(L)}:\ \|p\|+\|p\|^{1-\delta}>L\}\leq \# X_n^{(L)}-\# X_n^{(L-L^{1-\delta})}=\#X_n^{(L)}\cap (B_L\setminus B_{L-L^{1-\delta}})\leq C_2 L^{d-\delta},
      \]
      where the positive constant $C_2$ depends only on $r_0$ and $d$, since the $X_n$'s belong to $\mathcal{UD}_{r_0,d}$. Thus, we obtain that
      \begin{equation}\label{eq: perturproof2}
      \#\{p\in X_n^{(L)}:\ \|p+\xi_p^{(n)}\|>L\}\leq C_2 L^{d-\delta}+F_n.
      \end{equation}
      Hence, the first limit follows just by dividing by $L^d$ in \eqref{eq: perturproof2}, taking $\sup_{n\in\N}$ and by letting $L\to\infty$. The rest of the proof follows the very same lines as \cite{almostsure3}. 
\end{proof}

Now we prove that a convergence as in Lemma \ref{lemma: convFT} occurs at the level of the perturbations.

\begin{prop}\label{lemma: convFTper}
    Let $X_n,\ n\in\N$, and $X$  be discrete sets in $\mathcal{Q}_{r_0,d}$ such that $0\not\in X$. Suppose that $\xi_p^{(n)},\ p\in X_n$, are iid with law $\xi_p^{(n)}\sim\xi^{(n)}$, and that $\xi_p,\ p\in X$, are iid with law $\xi_p\sim\xi$. Also, assume that there exist $\epsilon>0$ and $s>0$ such that the following conditions are fulfilled:
    \begin{enumerate}
        \item $\mathbb{E}(\|\xi\|^{d+\epsilon})<\infty$;
        \item for every $p\in X_n$ and $q\in X$ with $\|p-q\|<s$, we have that $\xi_{p}^{(n)}$ and $\xi_q$ are independent.
     \end{enumerate}
    Then if $X_n\xrightarrow{\rho_{\mathsf{stat}}}X\in\mathcal{Q}_{r_0,d}$ and $\xi^{(n)}\xrightarrow{\mathsf{law}}\xi$, the following holds almost surely: 
    \[
    (\forall x\in\R^d),\quad
    \mathcal{F}_{(X_n)_{\xi^{(n)}}}(x)\to\mathcal{F}_{X_{\xi}}(x).
    \]
\end{prop}
\begin{proof}
    In what follows, we use $\mathcal{F}$ to denote the classical Fourier transform of a measure. Firstly, note that the convergence $X_n\xrightarrow{\rho_{\mathsf{stat}}} X$ implies that \eqref{eq: uniffar} holds for a large-enough positive integer $n$.
    Secondly, it should be noted that if the sequence of the $\xi^{(n)}$'s converges to $\xi$ in law, then from hypothesis 1 (after taking a subsequence if it is necessary), we have that
    \[
    \sup_{n\in\N}\mathbb{E}(\|\xi^{(n)}\|^{d+\epsilon})<\infty.
    \]
    Let $n\in\N$ be fixed, and $\epsilon>0$ being arbitrary. For each $n\in\N$, let us denote 
    \[
    E_{n,L}:=\#\{p\in X_n^{(L)}:\ p+\xi_p^{(n)}\not\in B_L\}\quad\text{and}\quad F_{n,L}:=\#\{p\in X_n\setminus B_L:\ p+\xi_p^{(n)}\in B_L\}.
    \]
    From Lemma \ref{lem: controlperb} we have that almost surely, for every $x\in\R^d$ there exists $L_0>0$ depending only on $\epsilon$ such that: 
    \begin{equation}\label{eq: convFTper1}
        \begin{split}
        \frac{1}{L^d}\left|\mathcal{F}(\delta_{(X_n)_{\xi}^{(L)}})(x)-\mathcal{F}(\delta_{X_{\xi}^{(L)}})(x)\right| & = \frac{1}{L^d}\left|\sum_{p\in (X_n)_{\xi}^{(L)}}e^{-2\pi i\langle p,x\rangle}-\sum_{p\in X_{\xi}^{(L)}}e^{-2\pi i\langle p,x\rangle}\right|\\
                    & \leq\frac{1}{L^d}\left|\sum_{p\in X_n^{(L)}}e^{-2\pi i\langle p+\xi_p^{(n)},x\rangle}-\sum_{p\in X^{(L)}}e^{-2\pi i\langle p+\xi_p,x\rangle}\right|\\
                    &+\frac{E_{n,L}}{L^d}+\frac{F_{n,L}}{L^d}\\
                    & \leq \frac{1}{L^d}\sum_{j=1}^{k_L}\left|e^{-2\pi i\langle p_{j,n}+\xi_{p_{j,n}}^{(n)},x\rangle}-e^{-2\pi i\langle p_j+\xi_{p_j},x\rangle}\right|+\epsilon\\
                    &+2\cdot\frac{\# A_{X_n,X}^{(L,\sigma)}\cup A_{X,X_n}^{(L,\sigma)}}{L^d}\\
                    &\leq\frac{1}{L^d}\left|\sum_{j=1}^{k_L}e^{-2\pi i\langle p_{j,n},x\rangle}(e^{-2\pi i\langle\xi_{p_{j,n}}^{(n)},x\rangle}-e^{-2\pi i\langle\xi_{p_{j}},x\rangle})\right|\\
                    &+\frac{1}{L^d}\left|\sum_{j=1}^{k_L}e^{-2\pi i\langle \xi_{p_j},x\rangle}(e^{-2\pi i\langle p_j,x\rangle}-e^{-2\pi i\langle p_{j,n},x\rangle})\right|+\epsilon+2\sigma,
        \end{split}
    \end{equation}
   where the bound in the first inequality in \eqref{eq: convFTper1} follows by applying Lemma \ref{lem: controlperb} simultaneously for each $n\in\N$; moreover the term $2\sigma$ appears from $X_n\xrightarrow{\rho_{\mathsf{stat}}}X$. Thus, from the fact that $|e^{-2\pi i\langle\cdot,x\rangle}|=1$, we get that
   \begin{equation}\label{eq: convFTper2}
   \begin{split}
       \frac{1}{L^d}\left|\mathcal{F}(\delta_{(X_n)_{\xi}^{(L)}})(x)-\mathcal{F}(\delta_{X_{\xi}^{(L)}})(x)\right|&\leq\frac{1}{L^d}\sum_{j=1}^{k_L}|e^{-2\pi i\langle\xi_{p_{j,n}}^{(n)},x\rangle}-e^{-2\pi i\langle\xi_{p_{j}},x\rangle}|\\
       &+\frac{1}{L^d}\sum_{j=1}^{k_L}|e^{-2\pi i\langle p_j,x\rangle}-e^{-2\pi i\langle p_{j,n},x\rangle}|+\epsilon+2\sigma.
      \end{split} 
   \end{equation}
   Note that the second sum in \eqref{eq: convFTper2} can be bounded from above by $\epsilon\cdot k_L$, as we prove in \eqref{eq: unifexp}. Now, observe that for any $n\in\N$ large-enough, and every $j,l\in\{1,\ldots, k_L\},\ j\neq l$, by the hypothesis 2, the random variables
    \[
    \left|e^{-2\pi i\langle\xi_{p_{j,n}}^{(n)},x\rangle}-e^{-2\pi i\langle\xi_{p_j},x\rangle}\right|\quad\text{and}\quad\left|e^{-2\pi i\langle\xi_{p_{l,n}}^{(n)},x\rangle}-e^{-2\pi i\langle\xi_{p_l},x\rangle}\right|
    \]
    are iid, and that they are (deterministic) uniformly bounded in $x$; then, as an application of the strong law of large numbers, we get almost surely that:
    \begin{equation}\label{eq: llnconFT}
        \frac{1}{k_L}\sum_{j=1}^{k_L}\left|e^{-2\pi i\langle\xi_{p_{j,n}}^{(n)},x\rangle}-e^{-2\pi i\langle\xi_{p_j},x\rangle}\right|\xrightarrow[L\to\infty]{}\mathbb{E}\left(\left|e^{-2\pi i\langle\xi^{(n)},x\rangle}-e^{-2\pi i\langle\xi,x\rangle}\right|\right).
    \end{equation}
    Since for each $n\in\N$ we have that $X_n\in\mathcal{UD}_{r_0,d}$, then there exists a positive number $C_0=C_0(r_0,d)$ such that $\frac{k_L}{L^d}\leq C_0$ for every $n\in\N$ (note that $k_L$ could depend on $n$). Hence, from this fact and by taking $L\to\infty$ in \eqref{eq: convFTper1}, from \eqref{eq: llnconFT} we obtain, almost surely, that:
    \begin{equation}\label{eq: convFTper3}
     (\forall\lambda\in\R^d),\quad\left|\mathcal{F}_{(X_n)_{\xi}}(x)-\mathcal{F}_{X_{\xi}}(x)\right|\leq C_0\cdot \mathbb{E}\left(\left|e^{-2\pi i\langle\xi^{(n)},x\rangle}-e^{-2\pi i\langle\xi,x\rangle}\right|\right)+\epsilon+2\sigma.
    \end{equation}

    The conclusion of the Lemma follows from \eqref{eq: convFTper3} and from the fact that $(\xi^{(n)})_{n\in\N}$ converges to $\xi$ in law.
\end{proof}

\begin{proof}[Proof of Theorem \ref{thm: C}]
    It follows from Propositions \ref{lemma: convFT} and \ref{lemma: convFTper}, and the continuity of the common recovering function $\Phi$. 
\end{proof}
\Addresses
\end{document}